\documentclass[10pt,twoside]{article}
\usepackage {amsmath, amssymb, amsthm, mathrsfs, amsfonts, bbm, ctable,graphicx,color,epsfig,citesort} 

\usepackage[abs]{overpic}

\usepackage{fancyhdr}
\renewcommand{\subsectionmark}[1]{}
\newlength\Li \newlength\Lii
\newtheorem{thm}{Theorem}[section]

\newcommand{\Dp}[1]{\partial_{\rho_{#1}}p/\gamma_{#1}}
\newcommand{\indet}{Z}

\newcommand{\diag}{\mbox{{diag}}}
\newcommand{\Spec}{\mbox{{Spec}}}
\newcommand{\sgn}{\mbox{{sgn}}}

\date{}

\makeatletter
\def\@seccntformat#1{}
\makeatother
\numberwithin{equation}{section}
\renewcommand{\numberline}[1]{}

\makeatletter
\def\blfootnote{\xdef\@thefnmark{}\@footnotetext}
\makeatother

\title{A Discontinuous Galerkin Method for Viscous Compressible Multifluids} 
\author{C.~Michoski\textsuperscript{\dag},\ J.A.~Evans\textsuperscript{*},  \ P.G.~Schmitz\textsuperscript{\ddag} \ \& \ A.~Vasseur\textsuperscript{**} \\ \\ \small{Departments of Mathematics}, \\ \small{Computational and Applied Mathematics,} \\ \small{Chemistry and Biochemistry} \\ \small{University of Texas at Austin}}

\usepackage[square,comma,numbers,sort&compress]{natbib}
\usepackage[pdftex,bookmarks]{hyperref}
\usepackage{hypernat}
 
\begin{document}
\maketitle
\begin{abstract}
We present a generalized discontinuous Galerkin method for a multicomponent compressible barotropic Navier-Stokes system of equations.  The system presented has a functional viscosity $\nu$ which depends on the pressure $p=p(\rho,\mu_{i})$ of the flow, with the density $\rho$ and the local concentration $\mu_{i}$.  High order Runge-Kutta time discretization techniques are employed, and different methods of dealing with arbitrary coupled boundary conditions are discussed.   Analysis of the energy consistency of the scheme is performed in addition to inspection of the relative error of the solution compared to exact analytic test cases.  Finally several examples, comparisons, generalizations and physical applications are presented.  \\ \\ \small{{\bf Keywords}:  Navier-Stokes; discontinuous Galerkin; Runge-Kutta; mixing; barotropic; compressible; viscous; miscible; multicomponent; multiphase; multifluid; chemical; acoustic.}
\end{abstract}

\blfootnote{\textsuperscript\textdagger {\it michoski@cm.utexas.edu}, Department of Chemistry and Biochemistry}\blfootnote{*{\it evans@ices.utexas.edu}. Computational and Applied Mathematics} \blfootnote{\textsuperscript\ddag {\it pschmitz@math.utexas.edu}, Department of Mathematics}\blfootnote{**{\it vasseur@math.utexas.edu}, Department of Mathematics}

\tableofcontents

\section{\texorpdfstring{\protect\centering $\S 1$ Introduction}{\S 1 Introduction}}

Much work has been done in the study of the numerics of multicomponent flows.  An example of an early yet comprehensive study of computational multiphase mechanics was given by Harlow and Amsden in Ref.~\cite{HA}, where they developed an implicit finite differencing technique for extremely generalized multicomponent settings of both compressible and incompressible flows, including phenomena ranging from bubble formation and cavitation effects, to the formation of atmospheric precipitation and mixing jets.  Subsequent and related work in multicomponent flows followed with, for example, the work of J.~Dukowicz in Ref.~\cite{Dukowicz} for particle-fluid models of incompressible sprays, an approach extended by G.~Faeth in Ref.~\cite{Faeth} to combustion flows and by D.~Youngs in Ref.~\cite{Youngs} to interfacial turbulent type flows. 

Owing to some of these pioneering works, recent work has demonstrated a resurgence of interest in multicomponent flows, approaches and numerical techniques.  The importance of fluid-flows comprised of more than one phase, chemical constituent, species or component is represented by a vast array of applications that range across a number of fields.  For example, multicomponent flows are essential for any flow demonstrating even rudimentary chemical kinetics; hence, for all (nontrivial) ``chemical fluids'' \cite{Williams}.  Likewise biological flows often require phase separations, in order to resolve membrane dynamics and interfacial behaviors in cells and cell organelles \cite{SIG} and medical applications desire estimates in local componentwise variations in blood serosity, which effect the viscosity and flow parameters involved with pulsatile hemodynamics \cite{BRaja}.  Likewise we find numerous examples of multicomponent flow applications in the atmospheric \cite{HG} and geophysical \cite{Vallis} sciences; as well as in acoustics \cite{MS} and astrophysics \cite{OD}, just to mention a few. 

Here we present a new multicomponent numerical scheme based on a mathematically well-posed \cite{MV} compressible barotropic system with functional viscosity depending on both the density $\rho$ and the mass fraction $\mu_{i}$ of each fluid component.  It is well-known, both experimentally and theoretically, that viscosity has a functional relationship to the density and specie type (for examples see the NIST thermophysical properties server).  In addition, these types of mathematical models (with functional transport coefficients) are well understood from the point of view of continuum dynamics, having been extensively studied by M{\'a}lek, Rajagopal et al. in Ref.~\cite{MR1,MR2,FMR}.  It is further seen in Ref.~\cite{MV} that the analytic model used in this work \emph{a priori} satisfies two essential entropy inequalities, much like the shallow water equations \cite{BD2}, which serve as important tools for numerical analysis and implementation. 

In this paper we implement a discontinuous Galerkin (DG) finite element method, employing piecewise polynomial approximations which do not enforce or require any type of continuity between the interfaces of ``neighboring'' elements.  This particular implementation is primarily motivated by the works of Cockburn, Shu et al. (see Ref.~\cite{Cockburn1,Cockburn2,Cockburn3,Cockburn4,Cockburn5}) and Feistauer, Dolej{\v{s}}{\'{\i}} et al. (see Ref.~\cite{FFS,FK,FD1,FD3}).  We implement a generalized formulation that is designed to accommodate an arbitrary choice of inviscid, viscous, and supplementary numerical fluxes.  We use explicit time discretization methods as described in Ref.~\cite{Cockburn1}, which necessitate a conditional stability requirement; namely the time discretization must satisfy the CFL condition.  Up to the CFL stability condition we find our method to be very robust and to deal well with arbitrary numbers of fluid components of arbitrary type --- up to the additional assumption that a barotropic pressure law is applicable.  On the domain boundary data we again strive to generalize our setting.  We show two different implementations of boundary conditions, which demonstrate different solvency with respect to interior solutions, initial conditions and phenomenolgically relevant contexts.  In both cases arbitrary Robin type BCs may be set.

In \textsection{2} we give the general governing system of equations, the mathematical regularity, and the discrete formulation of the problem.  In \textsection{3} we demonstrate a general way of dealing with boundary conditions by way of the method of characteristics, or alternatively, by way of setting arbitrary $L^\infty$ data on the boundary.  We provide an explicit formulation of the characteristic technique and show the generalized behavior of these types of ``characteristic'' boundary conditions, while subsequently discussing a number of alternative approaches.  In \textsection{4} we implement two test cases with exact solutions, which are restrictions placed on the multifluid barotropic governing equations, and show that they are exact up to the possible exception of the boundary data.  In \textsection{5} we show an example of a bifluid solution using the forward Euler method.  We then show the difference between boundary conditions by way of \emph{weak entropy} solutions versus that of characteristic boundary solutions.  The next section, \textsection{6}, is used to generalize the setting to $n$-fluid components and $k$-th order Runge-Kutta schemes, where the example of an $\ell=5$ fluid is shown explicitly.  Then in \textsection{7} we analyse the energy consistency of the modelisation with respect to two entropy inequalities derived in Ref.~\cite{MV}; one the classical entropy $\mathscr{S}$ and the second a closely related entropy $\tilde{\mathscr{S}}$ discovered by Bresch and Desjardins (see Ref.~\cite{BD3,BD4}), where it turns out that the numerical scheme from \textsection{2} satisfies both energy relations provided the CFL condition is satisfied.  Finally in \textsection{8} we extend the results to include Fick's diffusion law, where we inspect the exotic physical setting of a pressure wave traveling through a gas comprised partially of polyynes, and discuss some applications.  

\section{\texorpdfstring{\protect\centering $\S 2$ The generalized $\ell$-fluid}{\S 2 The generalized l-fluid}}

We consider a one dimensional compressible barotropic $\ell$-fluid system governed by the following system of equations:
\begin{align}
\label{mass}&\partial_{t}\rho+\partial_{x}(\rho u)=0,\\
\label{momentum}&\partial_{t}(\rho u)+\partial_{x}(\rho u^{2})+\partial_{x}p -\partial_{x}(\nu\partial_{x}u)=0, \\
\label{specie}&\partial_{t}(\rho\mu_{i})+\partial_{x}(\rho u\mu_{i})=0,
\end{align}
with initial conditions,
\[
\rho_{|t=0}=\rho_{0}>0,\quad \rho u_{|t=0}=m_{0}, \quad (\rho\mu_{i})_{|t=0}=\rho_{i,0}.
\]
The multicomponent barotropic pressure $p=p(\rho\mu_{1},\ldots,\rho\mu_{n})$ is chosen to satisfy,
\begin{equation}
\label{pressure}
p=\sum_{i=1}^{\ell}(\rho\mu_{i})^{\gamma_{i}},
\end{equation}
where $\sum_{i=1}^{\ell}\mu_{i}=1$. The mass conservation (\ref{mass}), momentum conservation (\ref{momentum}), species conservation (\ref{specie}), and barotropic equation of state (\ref{pressure}) describe the flow of a barotropic compressible viscous fluid defined for $(t,x)\in \mathbb{R}^{+}\times\mathbb{R}$.  Here the \emph{density} is given as $\rho$, the \emph{velocity} as $u$, the \emph{momentum} by $m$, and the \emph{mass fraction} of each component (chemical specie, phase element, etc.) of the fluid is given by $\mu_{i}$, respectively, where $\gamma_{i}>1$ corresponds to the emperically determined adiabatic exponent uniquely characterizing each of the $\ell$ species.  Furthermore, adopting the notation throughout the paper that $\rho_{i}=\rho\mu_{i}$, the form of the viscosity functional $\nu=\nu(\rho_{1},\ldots ,\rho_{\ell})$ is fixed to satisfy
\begin{equation}
\label{viscosity}
\nu = \psi'(p)\sum_{i=1}^{\ell}\rho_{i}\partial_{\rho_{i}}p,
\end{equation}
for $\psi'(p)=Cp^{-\alpha}$ given $\alpha\in(0,1)$ and $C>0$ as emperically determined constants (see Ref.~\cite{MR1,MR2}) and \textsection{7}).  

The mathematical well-posedness of such a system (in the $\ell=2$ case) is given by the following theorem, which was proven by two of the authors in Ref.~\cite{MV}:
\begin{thm}
Given (\ref{pressure}) and (\ref{viscosity}) satisfying the conditions in Ref.~\cite{MV} with initial data $(\rho_{0}, u_{0},\mu_{0})$ satisfying
\[
\begin{aligned}
0 < \underline{\varrho}(0)\leq & \rho_{0} \leq \overline{\varrho}(0)< \infty, \\
\rho_{0}\in\dot{H}^{1}(\mathbb{R}), \quad u_{0} \in & H^{1}(\mathbb{R}), \quad \mu_{0} \in H^{1}(\mathbb{R}), \\
\int_{\Omega}\mathscr{E}(\rho_{0},& \mu_{0})dx< +\infty, \\
|\partial_{x}&\mu_{0}|\leq C\rho_{0},
\end{aligned}
\]
with $\overline{\varrho}(0)$, $\underline{\varrho}(0)$ positive constants and $\mathscr{E}_{0}$ the internal energy as given in Ref.~\cite{MV}, there exists a global strong solution to (\ref{mass})-(\ref{specie}) on $\mathbb{R}^{+}\times\mathbb{R}$ such that for every $T>0$ we have
\[
\begin{aligned}
\rho \in L^{\infty}(0,T; \dot{H}^{1}(\mathbb{R})), & \quad \partial_{t}\rho \in L^{2}((0,T)\times \mathbb{R}), \\
u \in L^{\infty}(0,T; H^{1} (\mathbb{R}))\cap L^{2} (0,T & ;H^{2}(\mathbb{R})), \ \partial_{t}u \in L^{2}((0,T)\times\mathbb{R}), \\
\mu_{x} \in L^{\infty}(0,T; L^{\infty}(\mathbb{R})), & \quad \partial_{t}\mu \in L^{\infty}(0,T;L^{2}(\mathbb{R})).
\end{aligned}
\]
Furthermore, there exist positive constants $\underline{\varrho}(T)$ and $\overline{\varrho}(T)$ depending only on $T$, such that
\[
0< \underline{\varrho}(T) \leq \rho(t,x) \leq \overline{\varrho}(T)< \infty, \quad\quad \forall (t,x)\in (0,T)\times\mathbb{R}.
\]
Additionally, when $\psi''(p)$, $\partial_{\rho\rho}p(\rho,\mu)$, and $\partial_{\rho\mu}p(\rho,\mu)$ are each locally bounded then this solution is unique. 
\end{thm}

Now notice that for an $\ell$-fluid written with respect to conservation variables, the state vector $\boldsymbol{U}$ can be written as the transpose of the $1\times m$ row vector
\[
\boldsymbol{U}= (\rho, \rho u,\rho_{1},\ldots \rho_{\ell})^{T},
\]
where $m=\ell+2$ characterizes the degrees of freedom of our chosen system of equations.  Note that we make this choice of a state vector for the sake of flexibility of representation and implementation (see for example \textsection{8}), where the strict degrees of freedom of the system (\ref{mass})-(\ref{specie}), due to the multiplicity of (\ref{mass}) in the conservation form of (\ref{specie}), is just $\ell+1$.  Nevertheless, consistent with our choice of an $(\ell+2)\times 1$ state vector $\boldsymbol{U}$, we obtain that the $m\times 1$ inviscid flux vector $\boldsymbol{f}$ satisfies 
\[
\boldsymbol{f}(\boldsymbol{U})=(\rho u, \rho u^{2}+p, \rho_{1} u,\ldots , \rho_{\ell}u)^{T},
\]
while the $m\times 1$ viscous flux $\boldsymbol{g}$ is given by 
\[
\boldsymbol{g}(\boldsymbol{U},\boldsymbol{U}_{x}) = (0,\nu u_{x}, 0, \ldots , 0)^{T}.
\]
In this notation (\ref{mass})-(\ref{specie}) can be expressed as
\begin{equation}
\label{system}
\boldsymbol{U}_{t} + \boldsymbol{f}_{x} = \boldsymbol{g}_{x},
\end{equation}
where the notation $(\cdot)_{\iota}$ corresponds to component-wise derivations with respect to the variable $\iota$.

The Jacobian matrix of the inviscid flux $J_{\boldsymbol{U}}\boldsymbol{f}(\boldsymbol{U})=\boldsymbol{\Gamma}(\boldsymbol{U})$ can be written as the $m\times m$ matrix:
\begin{equation}
\label{jac}
\boldsymbol{\Gamma}(\boldsymbol{U}) = \left( \begin{array}{cc|ccc}
   0    & 1  & 0 & \cdots & 0 \\
   -u^2 & 2u &  \partial_{\rho_{1}}p  & \cdots &  \partial_{\rho_{\ell}}p \\
   \hline
   -u\mu_1 & \mu_1 \\
   \vdots & \vdots & & u\mathbbm{I}_{\ell} \\
   -u\mu_n & \mu_n \\
   \end{array} \right),
\end{equation}
where $\mathbbm{I}_{\ell}$ is the $\ell \times \ell$ identity matrix.  An important feature of the barotropic pressure law (\ref{pressure}) is that it is not a homogeneous function in $\rho_{i}$, and thus the Jacobian $\boldsymbol{\Gamma}$ is \emph{not} formulated to satisfy $\boldsymbol{\Gamma}\boldsymbol{U}=\boldsymbol{f}$.  This contrasts, for example, with the compressible Navier-Stokes equations when using the monofluid form of the ideal gas law $p=R\rho\vartheta$ (see Ref.~\cite{FFS}).  It should be noted that some numerical fluxes and schemes are designed or derived by specifically exploiting this homogeneity with respect to the Jacobian matrix of the flux function (for example, see the Vijayasundaram flux as used in Ref.~\cite{FK,FFS}). Nevertheless, our numerical fluxes will be defined independently of the homogeneity property of the corresponding map, where $\Gamma$ simply satisfies $\boldsymbol{f}_{x}=\boldsymbol{\Gamma}\boldsymbol{U}_{x}$.

For the viscous flux $\boldsymbol{g}$ we define the $m\times m$ matrix,  \begin{equation}\label{viscmat}
\mathscr{K}(\boldsymbol{U}) = \partial_{\boldsymbol{U}_{x}}\boldsymbol{g}(\boldsymbol{U},\boldsymbol{U}_{x})=\nu \left( \begin{array}{cc|c}
   0    & 0  & \mathbf{0} \\
    -\frac{u}{\rho} & \frac{1}{\rho} & \mathbf{0} \\
   \hline
   \mathbf{0} & \mathbf{0} & \mathbf{0}\\
   \end{array} \right),
\end{equation} where here and below the $\mathbf{0}$'s are zero matrices of the appropriate sizes.  Clearly then (\ref{system}) satisfies \begin{equation}\label{matform0}\boldsymbol{U}_{t}+\boldsymbol{\Gamma}\boldsymbol{U}_{x}-(\mathscr{K}\boldsymbol{U}_{x})_{x}=0.\end{equation}  Further let us introduce the auxilliary function $\boldsymbol{\Sigma}$ such that we are concerned with solving the coupled system: \begin{equation}\begin{aligned}\label{matform2}\boldsymbol{U}_{t}+\boldsymbol{\Gamma}\boldsymbol{U}_{x}-(\mathscr{K}\boldsymbol{\Sigma})_{x}&=0 \\ \boldsymbol{\Sigma} - \boldsymbol{U}_{x}=0.\end{aligned}\end{equation}  The above equations comprise a first order system which can be effectively discretized using the DG method.

\begin{figure}
{\setlength{\unitlength}{4144sp}%

{\begin{picture}(4043,1500)(-800,-400)
{\thinlines
\put(  1,839){\circle*{90}}}%
{\put(  1,839){\circle*{90}}}%
{\put(586,839){\circle*{90}}}%
{\put(1126,839){\circle*{90}}}%
{\put(2296,839){\circle*{90}}}%
{\put(2746,839){\circle*{90}}}%
{\put(3556,839){\circle*{90}}}%
{\put(  1,839){\circle*{90}}}%
{\put(  1,289){\circle{90}}}%
{\put(586,289){\circle{90}}}%
{\put(1126,289){\circle{90}}}%
{\put(2296,289){\circle{90}}}%
{\put(2746,289){\circle{90}}}%
{\put(3556,289){\circle{90}}}%
{\put(586,-151){\circle*{90}}}%
{\put(1126,-151){\circle*{90}}}%
{\put(2296,-151){\circle*{90}}}%
{\put(2746,-151){\circle*{90}}}%
{\put(  1,839){\line( 1, 0){1350}}}%
{\put(3556,839){\line(-1, 0){1485}}}%
{\put( 46,299){\line( 1, 0){495}}}%
{\put(631,299){\line( 1, 0){450}}}%
{\put(1171,299){\line( 1, 0){180}}}%
{\put(2071,299){\line( 1, 0){180}}}%
{\put(2341,299){\line( 1, 0){360}}}%
{\put(2791,299){\line( 1, 0){720}}}%
\put(-44,929){$a$}
\put(3516,929){$b$}
\put(-89,569){$x_0$}
\put(541,569){$x_1$}
\put(1081,569){$x_2$}
\put(1590,279){$\ldots$}
\put(1590,819){$\ldots$}
\put(1590,-171){$\ldots$}
\put(2211,569){$x_{ne-2}$}
\put(2701,569){$x_{ne-1}$}
\put(3501,569){$x_{ne}$}
\put(181,119){$\mathcal{G}_1$}
\put(721,119){$\mathcal{G}_2$}
\put(3016,119){$\mathcal{G}_{ne}$}
\put(2386,119){$\mathcal{G}_{ne-1}$}
\put(451,-421){$\mathcal{K}_{12}$}
\put(991,-421){$\mathcal{K}_{23}$}
\put(2161,-501){$\mathcal{K}_{ne-2,ne-1}$}
\put(2611,-381){$\mathcal{K}_{ne-1,ne}$}
\end{picture}}}
\caption{The discretization of $\Omega$, distinguishing nodes, elements and neighbors, with boundary $\partial\Omega = \{a,b\}$.}
\label{fig:scheme}
\end{figure}
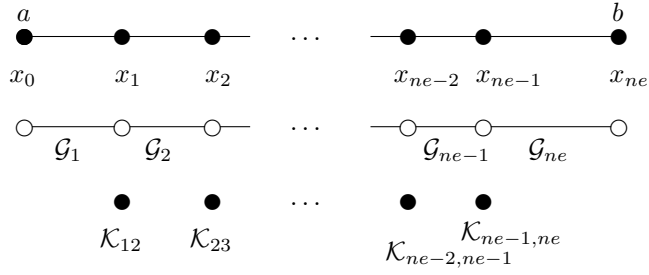

Consider the following discretization scheme motivated by Ref.~\cite{FFS} (and illustrated in the one dimensional case in Figure \ref{fig:scheme}).  Take an open $\Omega\subset\mathbb{R}$ with boundary $\partial\Omega=\Gamma$, given $T>0$ such that $\mathcal{Q}_{T}=((0,T)\times\Omega)$ for $\hat{\Omega}$ the closure of $\Omega$.  Let $\mathscr{T}_{h}$ denote the partition of the closure $\Omega$, such that taking $\hat{\Omega}=[a,b]$ provides the partition \[a=x_{0}<x_{1}\ldots<x_{ne}=b\] comprised of elements $\mathcal{G}_{i}=(x_{i-1},x_{i})\in\mathscr{T}_{h}$ such that $\mathscr{T}_{h}= \{\mathcal{G}_{1},\mathcal{G}_{2}, \ldots,\mathcal{G}_{ne}\}$.  The mesh diameter $h$ is given by $h=\sup_{\mathcal{G}\in\mathscr{T}_{h}}(x_{i}-x_{i-1})$ such that a discrete approximation to $\Omega$ is given by the set $\Omega_{h} = \cup_{i}\mathcal{G}_{i}\setminus\{a,b\}$.  Each element of the partition has a boundary set given by $\partial\mathcal{G}_{i} = \{x_{i-1},x_{i}\}$, where elements sharing a boundary point $\partial\mathcal{G}_{i}\cap\partial\mathcal{G}_{j}\neq\emptyset$ are characterized as neighbors and generate the set $\mathcal{K}_{ij}=\partial\mathcal{G}_{i}\cap\partial\mathcal{G}_{j}$ of interfaces between neighboring elements.  The boundary $\partial\Omega =\{a,b\}$ is characterized in the mesh as $\partial \Omega=\{x_{0},x_{ne}\}$ and indexed by elements $B_{j}\in\partial\Omega$ such that $\hat{\Omega} =\mathscr{T}_{h}\cup\mathcal{K}_{ij}\cup\partial\Omega$.  Now for $I\subset\mathbb{Z}^{+}=\{1,2,\ldots\}$ define the indexing set $r(i)=\{j \in I : \mathcal{G}_{j}$ is a neighbor of $\mathcal{G}_{i}\}$, and for $I_{B}\subset \mathbb{Z}^{-}=\{-1,-2,\ldots\}$ define $s(i)=\{j\in I_{B}:\mathcal{G}_{i}$ contains $B_{j}\}$.  Then for $S_{i}=r(i)\cup s(i)$, we have $\partial\mathcal{G}_{i}=\cup_{j\in S(i)}\mathcal{K}_{ij}$ and $\partial\mathcal{G}_{i}\cap\partial\Omega = \cup_{j\in s(i)}\mathcal{K}_{ij}$.     
    
We define the broken Sobolev space over the partition $\mathscr{T}_{h}$ as \[W^{k,2}(\Omega_{h},\mathscr{T}_{h})=\{v : v_{|\mathcal{G}_{i}}\in W^{k,2}(\mathcal{G}_{i}) \ \ \forall\mathcal{G}_{i}\in\mathscr{T}_{h}\}.\]  Further, approximate solutions to (\ref{mass})-(\ref{specie}) will exist in the space of discontinuous piecewise polynomial functions over $\Omega$ restricted to $\mathscr{T}_{h}$, given as \[S_{h}^{d}(\Omega_{h},\mathscr{T}_{h})=\{v:v_{|\mathcal{G}_{i}}\in \mathscr{P}^{d}(\mathcal{G}_{i}) \ \ \forall\mathcal{G}_{i}\in\mathscr{T}_{h}\}\] for $\mathscr{P}^{d}(\mathcal{G}_{i})$ the space of degree $\leq d$ polynomials on $\mathcal{G}_{i}$.  

Choosing a degree $d$ set of polynomial basis functions $N_{l}\in\mathscr{P}^{d}(\mathcal{G}_{i})$ for $l =0,\ldots, d$ we can denote an approximation to the state vector at the time $t$ over $\Omega_{h}$, by \[\boldsymbol{U}_{h}(t,x)=\sum_{l=0}^{d}\boldsymbol{U}_{l}^{i}(t)N^{i}_{l}(x),\quad  \forall x\in\mathcal{G}_{i},\] where the $N^{i}_{l}$'s are the finite element shape functions, and the $\boldsymbol{U}_{l}^{i}$'s correspond to the nodal unknowns.   Likewise the test functions $\boldsymbol{\varphi}_{h},\boldsymbol{\vartheta}_{h}\in W^{2,2}(\Omega_{h},\mathscr{T}_{h})$ are characterized by \[\begin{aligned}\boldsymbol{\varphi}_{h}(x)=\sum_{l=0}^{d}\boldsymbol{\varphi}_{l}^{i}N_{l}^{i}(x) \quad\mathrm{and}\quad \boldsymbol{\vartheta}_{h}(x)=\sum_{l=0}^{d}\boldsymbol{\vartheta}_{l}^{i}N_{l}^{i}(x)\quad \forall x\in\mathcal{G}_{i},\end{aligned}\] for $\boldsymbol{\varphi}_{l}^{i}$ and $\boldsymbol{\vartheta}_{l}^{i}$ the nodal values of the test function in each $\mathcal{G}_{i}$.   

Letting $\boldsymbol{U}$ be a classical solution to (\ref{matform2}) and multiplying through by test functions $\boldsymbol{\varphi}_{h}$ and $\boldsymbol{\vartheta}_{h}$ and integrating elementwise by parts yields:
\begin{equation}
\begin{aligned}
\label{weak}
\frac{d}{dt}\int_{\mathcal{G}_{i}}\boldsymbol{U}\cdot\boldsymbol{\varphi}_{h} dx + \int_{\mathcal{G}_{i}}(\boldsymbol{f}\cdot\boldsymbol{\varphi}_{h})_{x} dx - \int_{\mathcal{G}_{i}} \boldsymbol{f}\cdot\boldsymbol{\varphi}^{h}_{x} dx -\int_{\mathcal{G}_{i}} \boldsymbol{g}_{x}\cdot\boldsymbol{\varphi}_{h} dx = 0, \\ \int_{\mathcal{G}_{i}}\boldsymbol{\Sigma}\cdot \boldsymbol{\vartheta}_{h} dx - \int_{\mathcal{G}_{i}}(\boldsymbol{U}\cdot\boldsymbol{\vartheta}_{h})_{x} dx +\int_{\mathcal{G}_{i}}\boldsymbol{U}\cdot\boldsymbol{\vartheta}^{h}_{x}dx=0.
\end{aligned}
\end{equation}
 
Let $\varphi_{|\mathcal{K}_{ij}}$ and  $\varphi_{|\mathcal{K}_{ji}}$ denote the values of $\varphi$ on $\mathcal{K}_{ij}$ considered from the interior and the exterior of $\mathcal{G}_{i}$, respectivel. It should be noted that for $\mathcal{K}_{ij}\in\Gamma$, the restricted functions $\boldsymbol{\varphi}_{h}|\mathcal{K}_{ji}$ are determined up to a choice of boundary condition, which we will discuss in more detail in \textsection{3}.   Then we approximate the first term of (\ref{weak}) by, \begin{equation}
\begin{aligned}\label{timeterm}
\frac{d}{dt}\int_{\mathcal{G}_{i}}\boldsymbol{U}_{h}\cdot \boldsymbol{\varphi}_{h}dx  \approx  \frac{d}{dt}\int_{\mathcal{G}_{i}}\boldsymbol{U}\cdot\boldsymbol{\varphi}_{h}dx,
\end{aligned}
\end{equation} the second term in (\ref{weak}) by the inviscid numerical flux $\boldsymbol{\Phi}_{i}$,  
\begin{equation}
\begin{aligned}\label{invflux}
\tilde{\boldsymbol{\Phi}}_{i}(\boldsymbol{U}_{h}|_{\mathcal{K}_{ij}},\boldsymbol{U}_{h}|_{\mathcal{K}_{ji}}, \boldsymbol{\varphi}_{h}) & = \sum_{j\in S(i)}\int_{\mathcal{K}_{ij}}\boldsymbol{\Phi}(\boldsymbol{U}_{h}|_{\mathcal{K}_{ij}},\boldsymbol{U}_{h}|_{\mathcal{K}_{ji}},n_{ij})\cdot\boldsymbol{\varphi}_{h}|_{\mathcal{K}_{ij}} d\mathcal{K} \\ & \approx  \sum_{j\in S(i)}\int_{\mathcal{K}_{ij}}\boldsymbol{f}\cdot n_{ij}\boldsymbol{\varphi}_{h}|_{\mathcal{K}_{ij}}d\mathcal{K},
\end{aligned}
\end{equation}
for $n_{ij}$ the unit outward pointing normal; and the third term on the left in (\ref{weak}) by,
\begin{equation}\label{third}
\boldsymbol{\Theta}_{i}(\boldsymbol{U}_{h},\boldsymbol{\varphi}_{h})= \int_{\mathcal{G}_{i}} \boldsymbol{f}_{h} \cdot (\boldsymbol{\varphi}_{h})_{x} dx \approx -\int_{\mathcal{G}_{i}} \boldsymbol{f} \cdot (\boldsymbol{\varphi}_{h})_{x} dx.
\end{equation}

The viscous term in (\ref{weak}) integrates by parts to give:
\begin{equation}
\begin{aligned}\label{visflux}
\int_{\mathcal{G}_{i}} \boldsymbol{g}_{x}\cdot\boldsymbol{\varphi}_{h} dx & = \int_{\mathcal{G}_{i}}(\boldsymbol{g}\cdot\boldsymbol{\varphi}_{h})_{x}dx -\int_{\mathcal{G}_{i}}\boldsymbol{g}\cdot\boldsymbol{\varphi}^{h}_{x}dx \\
& =  \int_{\mathcal{G}_{i}}(\mathscr{K}\boldsymbol{\Sigma}\cdot\boldsymbol{\varphi}_{h})_{x}dx -\int_{\mathcal{G}_{i}}\mathscr{K}\boldsymbol{\Sigma}\cdot\boldsymbol{\varphi}^{h}_{x}dx.
\end{aligned}
\end{equation}
We approximate the first term on the right in (\ref{visflux}) using a generalized viscous flux $\hat{\mathscr{G}}$ (see Ref.~\cite{ABCM} for a review of choices in the DG framework). We write here for the general viscous flux
\begin{equation}
\begin{aligned}
\label{viscous}
\mathscr{G}_{i}(\boldsymbol{\Sigma}_{h},\boldsymbol{U}_{h},\boldsymbol{\varphi}_{h}) & = \sum_{j\in S(i)}\int_{\mathcal{K}_{ij}}\hat{\mathscr{G}}(\boldsymbol{\Sigma}_{h}|_{\mathcal{K}_{ij}},\boldsymbol{\Sigma}_{h}|_{\mathcal{K}_{ji}}, \boldsymbol{U}_{h}|_{\mathcal{K}_{ij}}, \boldsymbol{U}_{h}|_{\mathcal{K}_{ji}}, n_{ij})\cdot\boldsymbol{\varphi}_{h}|_{\mathcal{K}_{ij}} d\mathcal{K} \\ & \approx  \sum_{j\in S(i)}\int_{\mathcal{K}_{ij}}\boldsymbol{g}\cdot n_{ij}\boldsymbol{\varphi}_{h}|_{\mathcal{K}_{ij}}d\mathcal{K},
\end{aligned}
\end{equation} 
 while the second term is approximated by:
\begin{equation}\label{second}\mathscr{N}_{i}(\boldsymbol{\Sigma}_{h},\boldsymbol{U}_{h},\boldsymbol{\varphi}_{h})=\int_{\mathcal{G}_{i}}\boldsymbol{g}_{h}\cdot(\boldsymbol{\varphi}_{h})_{x}dx\approx \int_{\mathcal{G}_{i}}\boldsymbol{g}\cdot\boldsymbol{\varphi}^{h}_{x}dx.\end{equation} 

Finally for the second equation in (\ref{matform2}) we expand it such that the approximate solution satisfies: \begin{equation} \begin{aligned}\label{penalty} \mathscr{Q}_{i}(\hat{\boldsymbol{U}},\boldsymbol{\Sigma}_{h},\boldsymbol{U}_{h},\boldsymbol{\vartheta}_{h},\boldsymbol{\vartheta}_{x}^{h}) & =\int_{\mathcal{G}_{i}} \boldsymbol{\Sigma}_{h}\cdot \boldsymbol{\vartheta}_{h}dx  + \int_{\mathcal{G}_{i}}\boldsymbol{U}_{h}\cdot \boldsymbol{\vartheta}^{h}_{x}dx \\ & - \sum_{j\in S(i)}\int_{\mathcal{K}_{ij}}\hat{\boldsymbol{U}}(\boldsymbol{U}_{h}|_{\mathcal{K}_{ij}},\boldsymbol{U}_{h}|_{\mathcal{K}_{ji}},\boldsymbol{\vartheta}_{h}|_{\mathcal{K}_{ij}}) d\mathcal{K},\end{aligned}\end{equation} where, \[\begin{aligned}\mathscr{U}_{i}(\boldsymbol{U}_{h},\boldsymbol{\vartheta}_{h}) & = \sum_{j\in S(i)}\int_{\mathcal{K}_{ij}}\hat{\boldsymbol{U}}(\boldsymbol{U}_{h}|_{\mathcal{K}_{ij}},\boldsymbol{U}_{h}|_{\mathcal{K}_{ji}},\boldsymbol{\vartheta}_{h}|_{\mathcal{K}_{ij}}) d\mathcal{K} \\ &  \approx \sum_{j\in S(i)}\int_{\mathcal{K}_{ij}}\boldsymbol{U}\cdot n_{ij} \boldsymbol{\vartheta}_{h}|_{\mathcal{K}_{ij}}d\mathcal{K}\end{aligned}\] given that $\hat{\boldsymbol{U}}$ is the generalized flux term associated with the discontinuous Galerkin method determined up to a congeries of options (please see Ref.~\cite{ABCM} for a unified analysis), and using the approximate relations: \[\int_{\mathcal{G}_{i}} \boldsymbol{\Sigma}_{h}\cdot \boldsymbol{\vartheta}_{h}dx \approx \int_{\mathcal{G}_{i}} \boldsymbol{\Sigma}\cdot \boldsymbol{\vartheta}_{h}dx, \quad\mathrm{and}\quad \int_{\mathcal{G}_{i}}\boldsymbol{U}_{h}\cdot \boldsymbol{\vartheta}^{h}_{x}dx\approx  \int_{\mathcal{G}_{i}}\boldsymbol{U}\cdot \boldsymbol{\vartheta}^{h}_{x}dx.\]
We note that these choices of approximations and fluxes define the values of $\boldsymbol{\Sigma}_{h}$ on each element in terms of the values of $\boldsymbol{U}_{h}$ on that element and adjacent elements.  As we shall see later, this indicates that with an explicit time-discretization scheme, computing $\boldsymbol{\Sigma}_{h}$ is a completely local procedure.

Combining (\ref{invflux}), (\ref{third}), (\ref{viscous}), (\ref{second}) and (\ref{penalty}) and setting, \[\mathscr{X} = \sum_{\mathcal{G}_{i}\in\mathscr{T}_{h}}\mathscr{X}_{i},\] given the inner product \[(\boldsymbol{a}_{h}^{n},\boldsymbol{b}_{h})_{\Omega_{\mathcal{G}}} = \sum_{\mathcal{G}_{i}\in\mathscr{T}_{h}}\int_{\mathcal{G}_{i}}\boldsymbol{a}_{h}^{n}\cdot\boldsymbol{b}_{h} dx,\] we define an approximate solution to (\ref{weak}) as functions $\boldsymbol{U}_{h}$ and $\boldsymbol{\Sigma}_{h}$ for all $t\in (0,T)$ satisfying:
\begin{equation}
\begin{aligned}
\label{aprox}
& 1) \ \boldsymbol{U}_{h}\in C^{1}([0,T]; S_{h}^{d}), \ \ \boldsymbol{\Sigma}_{h}\in S_{h}^{d}, \\
& 2) \ \frac{d}{dt}(\boldsymbol{U}_{h},\boldsymbol{\varphi}_{h})_{\Omega_{\mathcal{G}}}+\tilde{\boldsymbol{\Phi}}(\boldsymbol{U}_{h},\boldsymbol{\varphi}_{h}) -  \boldsymbol{\Theta}(\boldsymbol{U}_{h},\boldsymbol{\varphi}_{h}) \\ & \quad\qquad\qquad\qquad\qquad\qquad-\mathscr{G}(\boldsymbol{\Sigma}_{h},\boldsymbol{U}_{h},\boldsymbol{\varphi}_{h})+ \mathscr{N}(\boldsymbol{\Sigma}_{h},\boldsymbol{U}_{h},\boldsymbol{\varphi}_{h})=0, \\  & 3) \ \mathscr{Q}(\hat{\boldsymbol{U}},\boldsymbol{\Sigma}_{h},\boldsymbol{U}_{h},\boldsymbol{\vartheta}_{h},\boldsymbol{\vartheta}_{x}^{h}) = 0, \\
& 4) \ \boldsymbol{U}_{h}(0)=\boldsymbol{U}_{0}.
\end{aligned}
\end{equation} We find below that up to a (possibly arbitrary) choice of boundary data, these solutions are quite well-behaved, extremely robust for arbitrary choice of $n$ fluids (we show $n=1,2$ and $5$ here, and have tested up to $n=11$ elsewhere) and readily extended to more complicated systems (e.g. \textsection{8}).  The results presented in this paper utilize piecewise linear basis functions, but we have tested quadratic basis functions in our code as well.

\section{\texorpdfstring{\protect\centering $\S 3$ Towards a generalized boundary treatment}{\S 3 Towards a generalized boundary treatment}}

Specifying arbitrary boundary data with respect to our approximate solution (\ref{aprox}) is a delicate issue which requires a nuanced understanding of barotropic solutions and the mathematical techniques used to pose them.  That is, we wish to determine the nature of an arbitrary boundary state $\boldsymbol{U}_{|\partial\Omega}$ in a way which is well-posed with respect to the system (\ref{mass})-(\ref{specie}); which is to say, in such a way that the uniqueness of the solution is maintained. 
 
However, practically speaking, recovering boundary data of an arbitrary nature on $\partial\Omega$ poses well-established difficulties with respect to the \emph{a priori} estimates established in Ref.~\cite{MV}, which serve as the cornerstone to the existence and uniqueness result stated in Theorem 2.1.  That is, recovering the \emph{a priori} estimates on the solution is reduced, in the first step, to recovering two entropy inequlities (see \textsection{7} for explicit forms) which serve as positive definite functionals over $(0,T)\times\Omega$.  However, when explicit boundary data is given, these inequalities acquire the addition of the following two unsigned boundary components, respectively (see Ref.~\cite{MV} for the explicit calculation): \[\int_{\Omega}(\rho u^{3})_{x}dx \quad\quad \mathrm{and}\quad \quad \int_{\Omega}(\rho u (u +\rho^{-1}\psi_{x})^{2})_{x}dx,\] having the consequence of rendering the well-posedness of a formulation which spans any type of boundary data difficult to establish.  Instead we offer a number of pragmatic approximate approaches that generalize the solution up to important restrictions, and then discuss some alternative approaches that are aimed at certain specialized types of settings.  First we review some known results.


It has been shown by Strikwerda in Ref.~\cite{Strikwerda}, and Gustaf'sson and Sundstrom in Ref.~\cite{GS} that incompletely parabolic systems, such as the shallow water equations and the full Navier-Stokes equations, may be well-posed with respect to a broad set of initial-boundary data.  These works additionally demonstrate the appropriate number of boundary conditions expected on incompletely parabolic systems, which differ from completely hyperbolic systems such as Euler's equations.  As the barotropic system (\ref{mass})-(\ref{momentum}) maintains a formal equivalence to the viscous shallow water equations (see for example Ref.~\cite{MV3,BD2}), we might expect (\ref{mass})-(\ref{momentum}) to behave as an incompletely parabolic system due to Ref.~\cite{GS}.  However, the dependencies of the pressure $p$ and viscosity $\nu$ on the mass fractions make showing this nontrivial and require a careful analysis of either incompletely parabolic systems \cite{Strikwerda}, or hyperbolic-parabolic systems \cite{KKP}. 

The implementation of both incompletely parabolic and hyperbolic systems often rely upon the so-called ``characteristic treatment.''  In these systems we use characteristic directions to extrapolate values of the system variables on the boundary, while the others become constrained by a set of characteristic relations (see Ref.~\cite{Kriess} for the hyperbolic regime).  These types of treatments have been extended to treat the full Navier-Stokes equations \cite{PLele,RS}, the viscous shallow water equations \cite{Lie}, and multifluid systems \cite{SK}.  
 
We want to consider what we will refer to here and below as characteristic type boundary solutions, which we view as a reduced hyperbolic system (as presented in Ref.~\cite{FFS,GS}).   We illustrate the situation for a simple one dimension case, but our analysis easily extends to the multidimensional case.  To begin, suppose we have the domain $(0,\infty)$ in which to specify a characteristic boundary condition at the boundary point $x=0$ and time $t=0$.  Note that other one dimensional cases can easily be transformed to such a setting with a change of coordinates.  We linearize our solution at $x=0$ with respect to a reference solution $\tilde{\boldsymbol{q}}$, which for our purposes represents the numerical solution at $x=0$ taken at a previous timestep.  As an approximation, we neglect the viscous terms, resulting in:
\begin{equation}
\label{moc}
\frac{\partial\boldsymbol{q}}{\partial t} + \boldsymbol{\Gamma}(\tilde{\boldsymbol{q}})\frac{\partial\boldsymbol{q}}{\partial x}= 0,
\end{equation}
where $\boldsymbol{q} \approx \boldsymbol{U}$ is a linearized approximation to the exact solution.  Note that this arrives with a linear hyperbolic system.  We consider the initial-boundary value problem in the set $(0,\infty)\times(0,\infty)$ equipped with the initial condition
\begin{equation}
\boldsymbol{q}(0,x)=\tilde{\boldsymbol{q}}, \quad \mathrm{for}\quad x\in (0,\infty),
\end{equation}
and the boundary condition
\begin{equation}
\boldsymbol{q}(t,0)=\tilde{\boldsymbol{q}}_{b}(t). \quad \mathrm{for}\quad t\in (0,\infty),
\end{equation}
Our goal is to choose the boundary condition $\tilde{\boldsymbol{q}}_{b}(t)$ in such a way that the initial-boundary value problem is well-posed. To continue, we decompose into characteristic directions.  That is, note that since $\boldsymbol{\Gamma}$ is diagonalizable we have that $\boldsymbol{\Gamma}_{h}\boldsymbol{c}_{j}=\varsigma_{j}\boldsymbol{c}_{j}$, where the characteristic directions $\boldsymbol{c}_{j}$ are the eigenvectors of $\boldsymbol{\Gamma}$ associated to eigenvalues $\varsigma_{j}$ (see \textsection{4} for an example).  Then we can formulate the solution in the form
\begin{equation}
\label{soln}
\boldsymbol{q}(t,x)=\sum_{j=1}^{m}\lambda_{j}(t,x)\boldsymbol{c}_{j},
\end{equation}
where the initial and boundary data, respectively, satisfy
\begin{equation}
\label{solode}
\tilde{\boldsymbol{q}}=\sum_{j=1}^{m}\alpha_{j}\boldsymbol{c}_{j}, \quad \mathrm{and}\quad \tilde{\boldsymbol{q}}_{b}=\sum_{j=1}^{m}\beta_{j}\boldsymbol{c}_{j}.
\end{equation} 


It then follows (from Ref.~\cite{FFS} chapter 3, for example) that (\ref{moc}) can be written as $j$ initial-boundary value scalar problems:
\begin{equation}
\begin{aligned}
\label{inbo}
 &\frac{\partial\lambda_{j}}{\partial t} + \varsigma_{j}\frac{\partial\lambda_{j}}{\partial x} = 0 \quad \mathrm{in} \ (0,\infty)\times (0,\infty), \\
&\lambda_{j}(x,0)=\alpha_{j}, \quad\mathrm{for} \ x\in (0,\infty), \\
&\lambda_{j}(0,t)=\beta_{j} \quad\mathrm{for} \ t\in (0,\infty).
\end{aligned}
\end{equation}
The scalar problems (\ref{inbo}) may be solved via the method of characteristics, from which we obtain the solution,
\begin{equation}
\label{bcsoln}
\lambda_{j}(x,t)=
\begin{cases}
\alpha_{j}, &\quad \mathrm{for} \ x -t\varsigma_{j}< 0, \\
\beta_{j},  &\quad \mathrm{for} \ x - t\varsigma_{j} > 0,
\end{cases}
\end{equation}
which provides an explicit form to (\ref{soln}).  From (\ref{bcsoln}), we obtain the following conditions for the boundary data.
\begin{itemize}
\item If $\varsigma_{j} > 0$, then necessarily $\beta_{j}=\alpha_{j}$.  This is obtained by extrapolating the solution of $\lambda_{j}$ to the boundary $x=0$.
\item If $\varsigma_{j} = 0$, then $\beta_{j}$ may be freely chosen.  However, in some situations it may be useful to choose $\beta_{j}=\alpha_{j}$ for this case, such as an impermeable solid wall.
\item If $\varsigma_{j} < 0$, then $\beta_{j}$ may be freely chosen.
\end{itemize}

 Note that once we have selected well-posed characteristic boundary conditions, we utilize the transformation 
\begin{equation}\label{alpbet}
\tilde{\boldsymbol{q}}_{b} = \boldsymbol{V} (\beta_{1},\beta_{2},\beta_{3},\cdots,\beta_{m})^{T}
\end{equation}
to determine the consistent boundary conditions for the conservation variables. It turns out that for (\ref{mass})-(\ref{viscosity}) we can reduce this method to that of the essential choices listed in Table \ref{table:odesys}.  This corresponds with what we know of hyperbolic systems as shown in Ref.~\cite{FFS} and Ref.~\cite{Kriess}, with respect to the number of free and fixed conditions on the boundaries.  In Table \ref{table:odesys} we also include a number of physically motivated restrictions which should be taken into account when selecting our boundary conditions.


\begin{table}[t]
\caption{\emph{Choice of boundary conditions}}
\centering
\begin{tabular}{|c | c | c | c | c|}
\hline
\multicolumn{2}{|c|}{\bf{Boundary type}} & \bf{Restrictions} & \bf{Free} & \bf{Fixed} \rule{0pt}{3ex} \rule[-2ex]{0pt}{0pt} \\ 
\hline\hline
Subsonic & inlet & & $\beta_{2},\ldots , \beta_{m}$ & $\beta_{1}$  \rule{0pt}{3ex} \rule[-2ex]{0pt}{0pt} \\  $u\cdot \boldsymbol{n} > -c$ &  $u\cdot\boldsymbol{n}<0$ &  $\mu_{1}+\cdots+\mu_{n}=1$, $\rho>0$ & &  \rule{0pt}{2ex} \rule[-1ex]{0pt}{0pt} \\
\cline{1-2}\cline{4-5}
Supersonic & inlet & and the appropriate  & $\beta_{1},\ldots,\beta_{m}$, & none  \rule{0pt}{3ex} \rule[-2ex]{0pt}{0pt} \\   $u\cdot \boldsymbol{n} \leq -c$ & $u\cdot\boldsymbol{n}<0$ & supplimentary &   &  \rule{0pt}{2ex} \rule[-1ex]{0pt}{0pt} \\ 
\cline{1-2}\cline{4-5}
Subsonic & outlet &  conditions associated  & $\beta_{2}$ & $\beta_{1},\beta_{3},$  \rule{0pt}{3ex} \rule[-2ex]{0pt}{0pt} \\  $u\cdot \boldsymbol{n} < c$ &  $u\cdot\boldsymbol{n}> 0$ &  with a choice of &  & $\ldots,\beta_{m}$ \rule{0pt}{2ex} \rule[-1ex]{0pt}{0pt} \\
\cline{1-2}\cline{4-5}
Supersonic & outlet & boundary data, &  none & $\beta_{1},$  \rule{0pt}{3ex} \rule[-2ex]{0pt}{0pt} \\   $u\cdot \boldsymbol{n} \geq c$ &  $u\cdot\boldsymbol{n}>0$ &  including:  &  & $\ldots,\beta_{m}$  \rule{0pt}{2ex} \rule[-1ex]{0pt}{0pt} \\
\cline{1-2}\cline{4-5}
Membrane & wall  &  $\rho,u,\mu_{i},p,\nu,m,\rho\mu_{i}$, etc. &  $\beta_{2}$ &  $\beta_{1},\beta_{3},$  \rule{0pt}{3ex} \rule[-2ex]{0pt}{0pt} \\  $u\cdot\boldsymbol{n}= 0$ &  $u\cdot\boldsymbol{n}= 0$ &  &  & $\cdots, \beta_{m}$  \rule{0pt}{2ex} \rule[-1ex]{0pt}{0pt} \\
\cline{1-2}\cline{4-5}
Membrane & osmotic  &  &  $\beta_{2},\ldots , \beta_{m}$ &  $\beta_{1}$  \rule{0pt}{3ex} \rule[-2ex]{0pt}{0pt} \\  $u\cdot\boldsymbol{n}= 0$ &  $u\cdot\boldsymbol{n}= 0$ &  &  &  \rule{0pt}{2ex} \rule[-1ex]{0pt}{0pt} \\
\hline
\end{tabular}
\label{table:odesys}
\end{table}

In addition to employing these ``characteristic'' solutions, we notice that the form of (\ref{moc}) satisfies the \emph{weak entropy} solutions of Ref.~\cite{Cockburn1,BLN} and Ref.~\cite{Martin} for hyperbolic systems.  However, as we show in \textsection{4}, even though these two types of solutions are both consistent, they do not display equivalent numerical behavior.

Nevertheless these two choices of boundary data, the characteristic and weak entropy solutions, are not ideal since (\ref{mass})-(\ref{specie}) is \emph{not} a hyperbolic system.  We may alternatively consider the route of positing boundary data by a simple extension of the results of Zlotnik (see Ref.~\cite{Z2}) to see that the barotropic system is parabolic in the sense of Petrovskii upon addition of the ``quasihydrodynamic'' or ``quasigasdynamic'' auxilliary function $w$ (see Ref.~\cite{ES,Zhdanov}) on $\partial\Omega$.


More generally, there exists a large back catalogue of results on compressible barotropic systems, many of which implement differing initial-boundary data, and some of which utilize fairly exotic conditions on the boundary.  For example, for barotropic inflow problems we can refer to both Ref.~\cite{Kazenkin} and Ref.~\cite{MN}, where in both cases results from Ref.~\cite{MV} are required and additional extensions are needed to move into the multiphase regime.  Likewise solutions exist for free boundary barotropic problems \cite{ST}, surface tension type boundaries \cite{ST2}, Navier boundary type conditions \cite{BD7}, and various Dirichlet type problems near vacuum states \cite{MW,CCK,PLL2}; however, again, all of these results are only strictly satisfied for monofluidic systems, and thus require subtle analysis in order to extend them to the full multifluid regime. In many cases however, such as in Ref.~\cite{Z2}, the extension is relatively straightforward.
  
\section{\texorpdfstring{\protect\centering $\S 4$ Numerical test cases}{\S 4 Numerical test cases}}

We inspect two analytic test cases to verify the accuracy of the numerical method presented in \textsection{2} and \textsection{3}.  In both cases we solve a monofluid restriction of (\ref{mass})-(\ref{specie}) from the bifluid case ($\ell=2$), with $\mu_{1}=1$ and $\mu_{2}=0$ in $l=1$ spatial dimension.  

To begin, we specify the DG formulation in the bifluid case.  First we define the three vectors $\boldsymbol{U}= (\rho, \rho u,\rho_{1},\rho_{2})^{T}$, $\boldsymbol{f}(\boldsymbol{U})=(\rho u, \rho u^{2}+p, \rho_{1} u,\rho_{2} u)^{T}$ and $\boldsymbol{g}(\boldsymbol{U},\boldsymbol{U}_{x}) = (0,\nu u_{x},0)^{T}$ such that (\ref{mass})-(\ref{specie}) are expressed as \begin{equation}\label{system2}\boldsymbol{U}_{t} + \boldsymbol{f}_{x} = \boldsymbol{g}_{x},\end{equation} whereby setting $\ell=2$ in (\ref{jac}) and using (\ref{viscmat}) it then follows that \begin{equation}\boldsymbol{U}_{t} + \boldsymbol{\Gamma}\boldsymbol{U}_{x} = (\mathscr{K}\boldsymbol{U}_{x})_{x}.\end{equation}  We can thus write a weak form of (\ref{mass})-(\ref{specie}) in the same way as (\ref{weak}).

To solve the system we must first specify the inviscid flux $\boldsymbol{\Phi}$.  We test for two choices here.  First we implement the local Lax-Friedrich's flux $\boldsymbol{\Phi}_{lLF}$ which satisfies
\[
\begin{aligned}
\int_{\mathcal{K}_{ij}}\boldsymbol{\Phi}_{lLF}\cdot\boldsymbol{\varphi}_{h} d\mathcal{K} & =  \frac{1}{2}\int_{\mathcal{K}_{ij}}(\boldsymbol{f}(\boldsymbol{U}_{h})_{|\mathcal{K}_{ij}}+\boldsymbol{f}(\boldsymbol{U}_{h})_{|\mathcal{K}_{ji}})\cdot n_{ij}\boldsymbol{\varphi}_{h}|_{\mathcal{K}_{ij}}d\mathcal{K} \\
& - \frac{1}{2}\int_{\mathcal{K}_{ij}}(\Spec_{r}(\boldsymbol{\Gamma}))((\boldsymbol{U}_{h})_{|\mathcal{K}_{ij}}-(\boldsymbol{U}_{h})_{\mathcal{K}_{ji}})\cdot n_{ij}\boldsymbol{\varphi}_{h}|_{\mathcal{K}_{ij}} d\mathcal{K},
\end{aligned}
\]
for $n_{ij}$ the outward unit normal and $\Spec_{r}(\boldsymbol{\Gamma})$ the spectral radius of $\boldsymbol{\Gamma}$. 

As our second choice of inviscid flux we implement a standard approximate Riemann solver, with flux $\boldsymbol{\Phi}_{R}$ satisfying:
\[
\begin{aligned}
\int_{\mathcal{K}_{ij}}\boldsymbol{\Phi}_{R}\cdot\boldsymbol{\varphi}_{h} d\mathcal{K} & =  \frac{1}{2}\int_{\mathcal{K}_{ij}}(\boldsymbol{f}(\boldsymbol{U}_{h})_{|\mathcal{K}_{ij}}+(\boldsymbol{f}(\boldsymbol{U}_{h})_{|\mathcal{K}_{ji}})\cdot n_{ij}\boldsymbol{\varphi}_{h}|_{\mathcal{K}_{ij}}d\mathcal{K} \\
& - \frac{1}{2}\int_{\mathcal{K}_{ij}}(V(\{\boldsymbol{U}_{h}\})|\Lambda(\{\boldsymbol{U}_{h}\})|V^{-1}(\{\boldsymbol{U}_{h}\}))\cdot n_{ij}\boldsymbol{\varphi}_{h}|_{\mathcal{K}_{ij}} d\mathcal{K},
\end{aligned}
\]
where $V$ and $V^{-1}$ are found from the eigendecomposition given in the appendix, $\Lambda$ is given by the diagonal matrix of eigenvalues $\diag(\varsigma_{i})$ -- as also enumerated in the appendix -- and the average is given by $$\{\boldsymbol{U}_{h}\} = \frac{1}{2}\left(\boldsymbol{U}_{h}|_{\mathcal{K}_{ij}}+\boldsymbol{U}_{h}|_{\mathcal{K}_{ji}}\right).$$ 

Next we specify the viscous flux $\mathscr{G}$.  Here we use a formulation similar to that presented in Ref.~\cite{BR}, but we adapt it to include the functional dependencies present in the viscosity.  We choose \[\int_{\mathcal{K}_{ij}} \hat{\mathscr{G}}_{b}\cdot n_{ij}\boldsymbol{\varphi}_{h} d\mathcal{K} = \frac{1}{2}\int_{\mathcal{K}_{ij}}((\mathscr{K}\boldsymbol{\Sigma}_{h})_{|\mathcal{K}_{ij}} + (\mathscr{K}\boldsymbol{\Sigma}_{h})_{|\mathcal{K}_{ji}}) \cdot n_{ij}\boldsymbol{\varphi}_{h}|_{\mathcal{K}_{ij}} d\mathcal{K}.\]
For the numerical flux $\hat{\boldsymbol{U}}$ we use the Bassi-Rebay form, as shown in Ref.~\cite{BR,ABCM}, which gives \[\int_{\mathcal{K}_{ij}}\hat{\boldsymbol{U}}_{BR}(\boldsymbol{U}_{h},\boldsymbol{\vartheta}_{h})d\mathcal{K} = \frac{1}{2}\int_{\mathcal{K}_{ij}}((\boldsymbol{U}_{h})_{|\mathcal{K}_{ij}} +(\boldsymbol{U}_{h})_{|\mathcal{K}_{ji}})\cdot n_{ij}\boldsymbol{\vartheta}_{h}|_{\mathcal{K}_{ij}}d\mathcal{K}.\]

Now we discretize in time, denoting a partition of [0,T] by \[0=t^{0}<t^{1}\ldots<t^{T}=T,\] for a timestep given as $\Delta t^{n}=t^{n+1}-t^{n}$, and implement the forward Euler scheme:
\[ 
\frac{\partial\boldsymbol{U}_{h}}{\partial t} \approx \frac{\boldsymbol{U}_{h}^{n+1}-\boldsymbol{U}_{h}^{n}}{\Delta t^{n}}, 
\] along with a slope limiting scheme in the conservation variables $(\rho, \rho u)$, where the van Leer and Osher MUSCL schemes (as shown in Ref.~\cite{VL1,VL2,Osher}) have been adopted in this paper.


Now we solve explicitly for (\ref{aprox}).   In particular, we show an explicit scheme using the Riemann flux, which is formulated to read: for every $n\geq 0$ find $\boldsymbol{U}_{h}^{n+1}$ such that
\begin{equation}
\begin{aligned}
\label{approxtest}
& 1) \ \boldsymbol{U}^{n}_{h}\in S_{h}^{d}, \ \ \boldsymbol{\Sigma}_{h}^{n}\in S_{h}^{d}, \\
& 2) \  \left(\frac{\boldsymbol{U}_{h}^{n+1}-\boldsymbol{U}_{h}^{n}}{\Delta t^{n}}, \boldsymbol{\varphi}_{h}\right)_{\Omega_{\mathcal{G}}} +\tilde{\boldsymbol{\Phi}}_{R}(\boldsymbol{U}_{h}^{n},\boldsymbol{\varphi}_{h}) -  \boldsymbol{\Theta}(\boldsymbol{U}_{h}^{n},\boldsymbol{\varphi}_{h}) \\ & \quad\qquad\qquad\qquad\qquad\qquad-\mathscr{G}_{b}(\boldsymbol{\Sigma}_{h}^{n},\boldsymbol{U}_{h}^{n},\boldsymbol{\varphi}_{h})+ \mathscr{N}(\boldsymbol{\Sigma}_{h}^{n},\boldsymbol{U}_{h}^{n},\boldsymbol{\varphi}_{h})=0, \\  & 3) \ \mathscr{Q}(\hat{\boldsymbol{U}}_{BR},\boldsymbol{\Sigma}_{h}^{n},\boldsymbol{U}_{h}^{n},\boldsymbol{\vartheta}_{h},\boldsymbol{\vartheta}_{x}^{h}) = 0, \\
& 4) \ \boldsymbol{U}_{0}^{h} = \boldsymbol{U}_{h}(0).
\end{aligned}
\end{equation}
The above formulation lends itself naturally to a staggered scheme.  First, given $\boldsymbol{U}^{n}_{h}$ one solves step 3 for $\boldsymbol{\Sigma}^{n}_{h}$.  This amounts to a simple, fast, and trivially parallelizable computation as the $L^2$-projection matrix to be inverted is block-diagonal, with each block corresponding to an individual element.  Second, given $\boldsymbol{\Sigma}^{n}_{h}$, one solves step 2 for $\boldsymbol{U}^{n+1}_{h}$.  This similarly is a trivial computation as the mass matrix to be inverted is block-diagonal.  In fact, with the choice of an orthogonal polynomial basis on each element, the $L^{2}$-projection and mass matrices become diagonal.

\begin{figure}[t!]
\centering
\hspace{-8pt} \hspace{-8pt}\includegraphics[width=6.1cm]{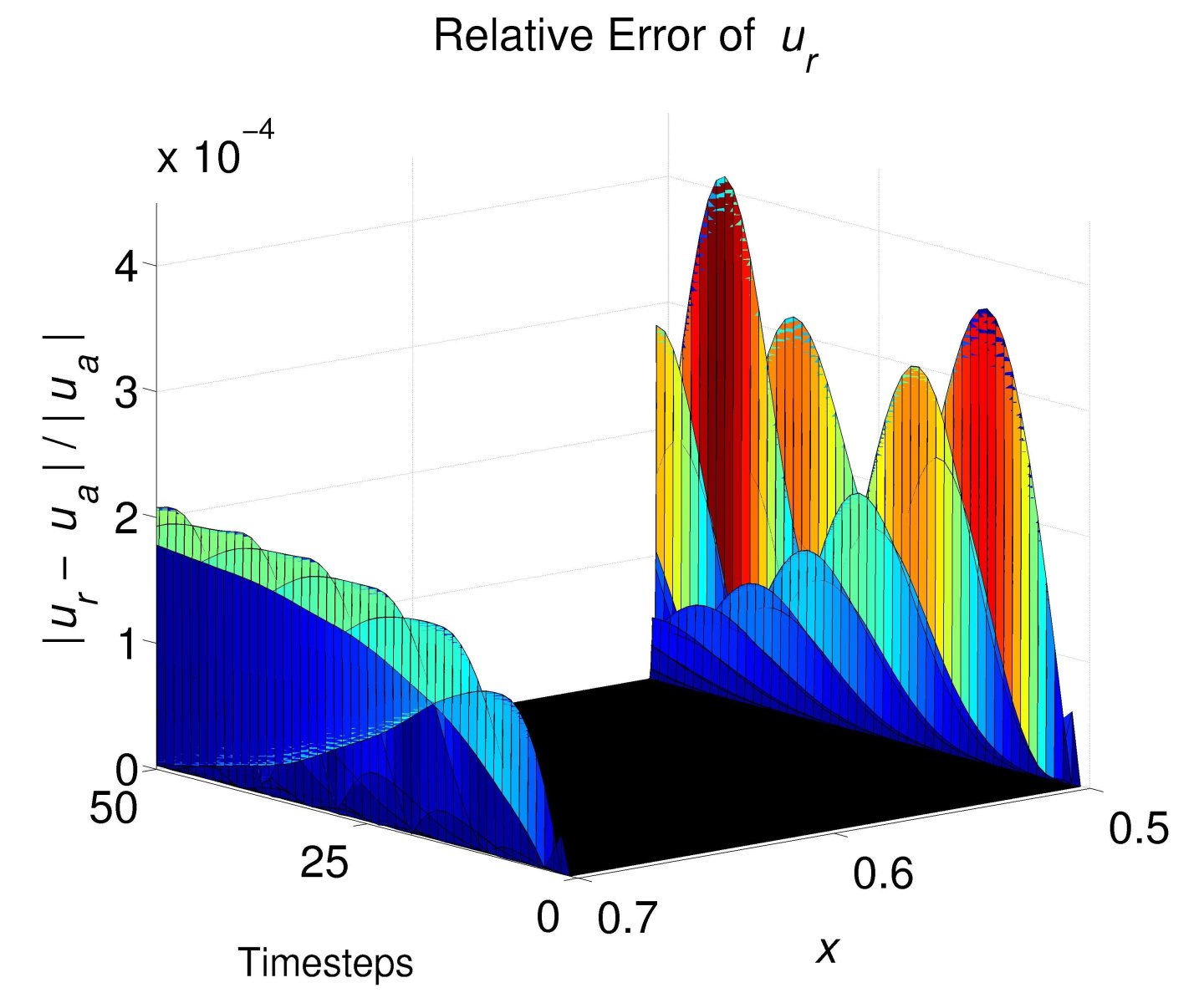}\hspace{-2pt}\includegraphics[width=6.1cm]{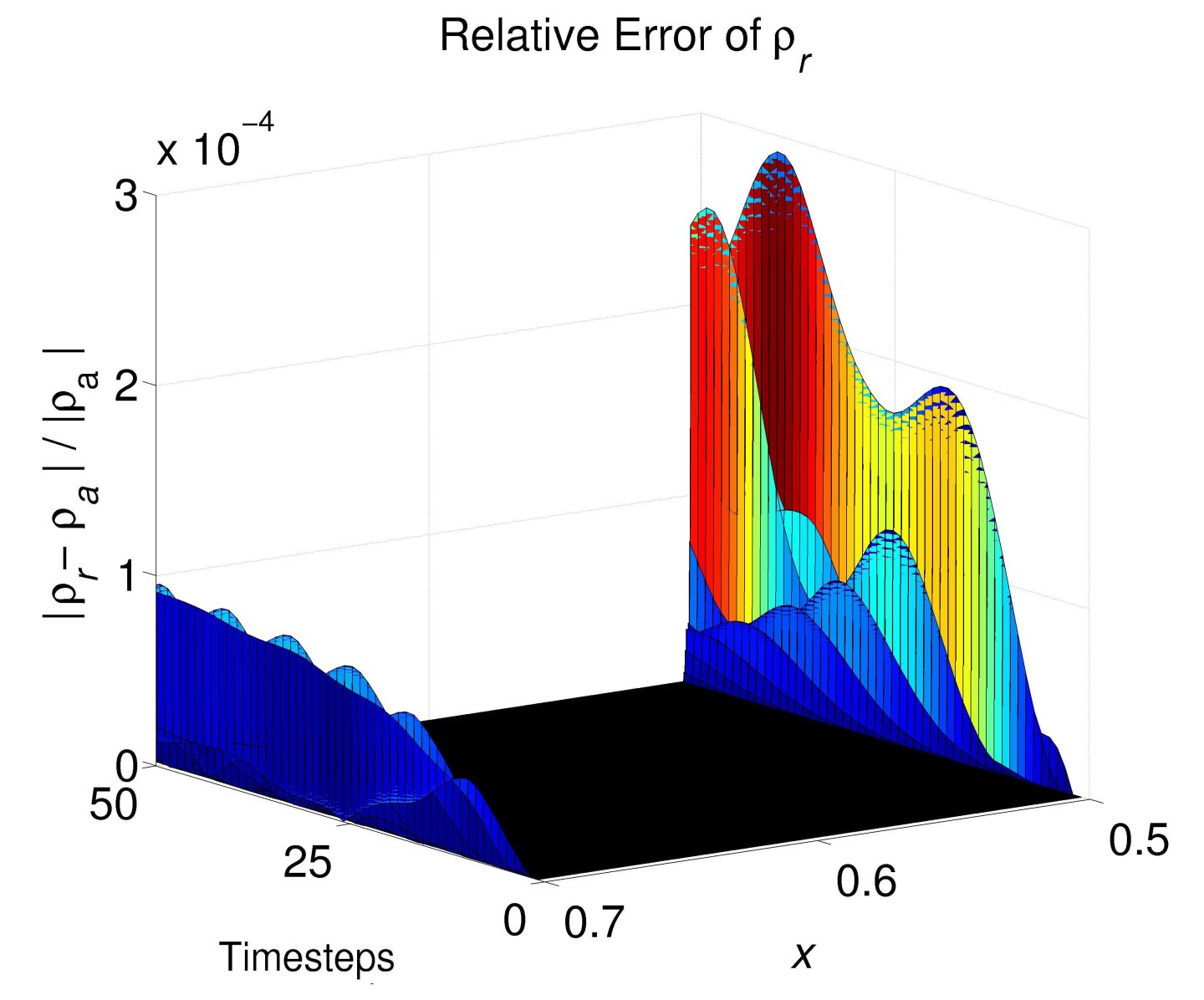}
\caption{ The two graphs show the solution to (\ref{uode}) in terms of the linear Riemann flux and the Osher limiter, denoted $u_{r}$ and $\rho_{r}$, versus the exact solution.}
\label{fig:ODE1}  
\end{figure}

We inspect the first of two numerical test cases.  Consider the monofluid steady state case of (\ref{mass})-(\ref{specie}), by setting the initial data to $\rho_{0} = \mu_{1,0}^{-1} = u_{0} = \gamma_{i} = 1$ and $\mu_{2,0}=0$.  Clearly here the pressure reduces to unity $p_{0}=1$ and the viscosity to a constant $\nu_{0} = C_{0}$.  Next we set the periodic boundary condition \[
\boldsymbol{U}_{h}^{n}(a^{+},t)=\boldsymbol{U}_{h}^{n}(b^{+},t), \hspace{10pt} \boldsymbol{U}_{h}^{n}(a^{-},t)=\boldsymbol{U}_{h}^{n}(b^{-},t) .
\]  The exact solution shows constant solutions in the primitive variables.  Our numerical simulations for \ref{approxtest} using both approximate Riemann and Lax-Friedrich's inviscid fluxes have shown that the $L^{\infty}$ numerical error in the conservation variables is of the order of machine precision, showing no fluctuation about the steady state in time.



For the second of our test cases, we consider the monofluidic restriction of (\ref{mass})-(\ref{specie}) given by taking $\mu_{1,0}=1$, $\mu_{2,0}=0$ and $\gamma_{i}=1$ with the additional relations: \[p=\rho = u^{-1},\qquad\mathrm{and}\qquad \nu = \rho.\]  Solving this system immediately yields \[\rho^{-1} + \rho - \rho \partial_{x}\rho^{-1} = -C,\] for $C\in\mathbb{R}$, which leads to the ordinary differential equation \begin{equation}\label{uode}u_{x} = u^{2} +1 - Cu.\end{equation}  Setting $C=0$ the noting that solution is independent of time, we solve the ODE yielding: $u =\tan{x}$.  Setting the initial data to $$\rho_{0}=(\tan{x})^{-1},\qquad m_{0} = 1,\qquad\mathrm{and}\qquad (\rho\mu_{1})_{0} =\rho_{0},$$ with the Dirichlet boundary data provided in the weak entropy sense of \textsection{3} via, $$\rho_{b} = 1/u_{b}, \qquad m_{b} = 1 \qquad \mathrm{and}\qquad (\rho\mu_{1})_{b} = \rho_{b},$$ we inspect the solution over the domain $[a,b]$, with $a=0.5$ and $b=0.7.$  Note that we enforce the weak entropy boundary conditions by setting $\boldsymbol{U}_h^{n}|_{K_{ji}}=\{ \rho_b, m_b, \rho_b, 0 \}$ in our discontinuous Galerkin formulation.  Here we compare the exact solution to the solution obtained using the Riemann flux with the Osher slope limiter (denoted $\rho_{r}$ in Figure \ref{fig:ODE1}).   

In Figure \ref{fig:ODE1} we plot the error of the numerical solution corresponding to a mesh size of $h=2\times 10^{-4}$ and a timestep size of $h/30$, where it is clear that the relative error over fifty timesteps is of the order of magnitude of the resolution of the mesh.  The relative error is zero across the solution at the first timestep, as expected, and remains nearly constant in the interior of the domain in both cases, while the weak entropy implementation displays fluctuations in time of the order of $h$. These boundary fluctuations are neither monotonic nor generally increasing, but show complicated temporal perturbations at the weak entropy boundary points and are seen to weakly propagate into the interior as a function of the timestep.  We have obtained similar behavior for the choices of a local Lax-Friedrich's inviscid flux and van Leer's slope limiter. Further, numerical experiments have revealed that the $L^{2}$-error of the solution at a fixed time $T$ scales like $O(h)$ for the choice of a backward Euler scheme, a timestep size of $\Delta t = h/30$, and piecewise linear basis functions.  For a general polynomial order $d$ and an explicit time integration scheme of order $k$ (see \textsection{6}), we find the $L^{2}$-error of the solution at a fixed time $T$ scales like $O(h^{d+1}+\Delta t^{k})$, as expected, provided the CFL condition is satisfied. 
   
\section{\texorpdfstring{\protect\centering $\S 5$ Example: $2$-fluid with chemical inlet}{\S 5 Example: 2-fluid with chemical inlet}}

Let us show a simple application of the system outlined in \textsection{2} and \textsection{3} evaluated over two distinct constituents.  Consider the bifluid system,
\begin{align}
\label{mass2}&\partial_{t}\rho+\partial_{x}(\rho u)=0,\\
\label{momentum2}&\partial_{t}(\rho u)+\partial_{x}(\rho u^{2})+\partial_{x}p -\partial_{x}(\nu\partial_{x}u)=0, \\
\label{specie2}&\partial_{t}(\rho\mu_{i})+\partial_{x}(\rho u\mu_{i})=0,
\end{align}
with initial conditions:
\[
\rho_{|t=0}=\rho_{0}>0,\qquad m_{|t=0}=m_{0} \qquad \mathrm{and} \qquad \mu_{t=0}=\mu_{0}.
\]
The pressure is given by $p=\rho_{1}^{\gamma_{1}}+\rho_{2}^{\gamma_{2}}$ and the viscosity by $\nu=\psi'(\gamma_{1}\rho_{1}^{\gamma_{1}}+\gamma_{2}\rho_{2}^{\gamma_{2}})$ for $\psi'=Cp^{-\alpha}$ and $\alpha\in(0,1)$ with $C>0$.

\begin{figure}[t!]
\centering
\includegraphics[width=6.2cm]{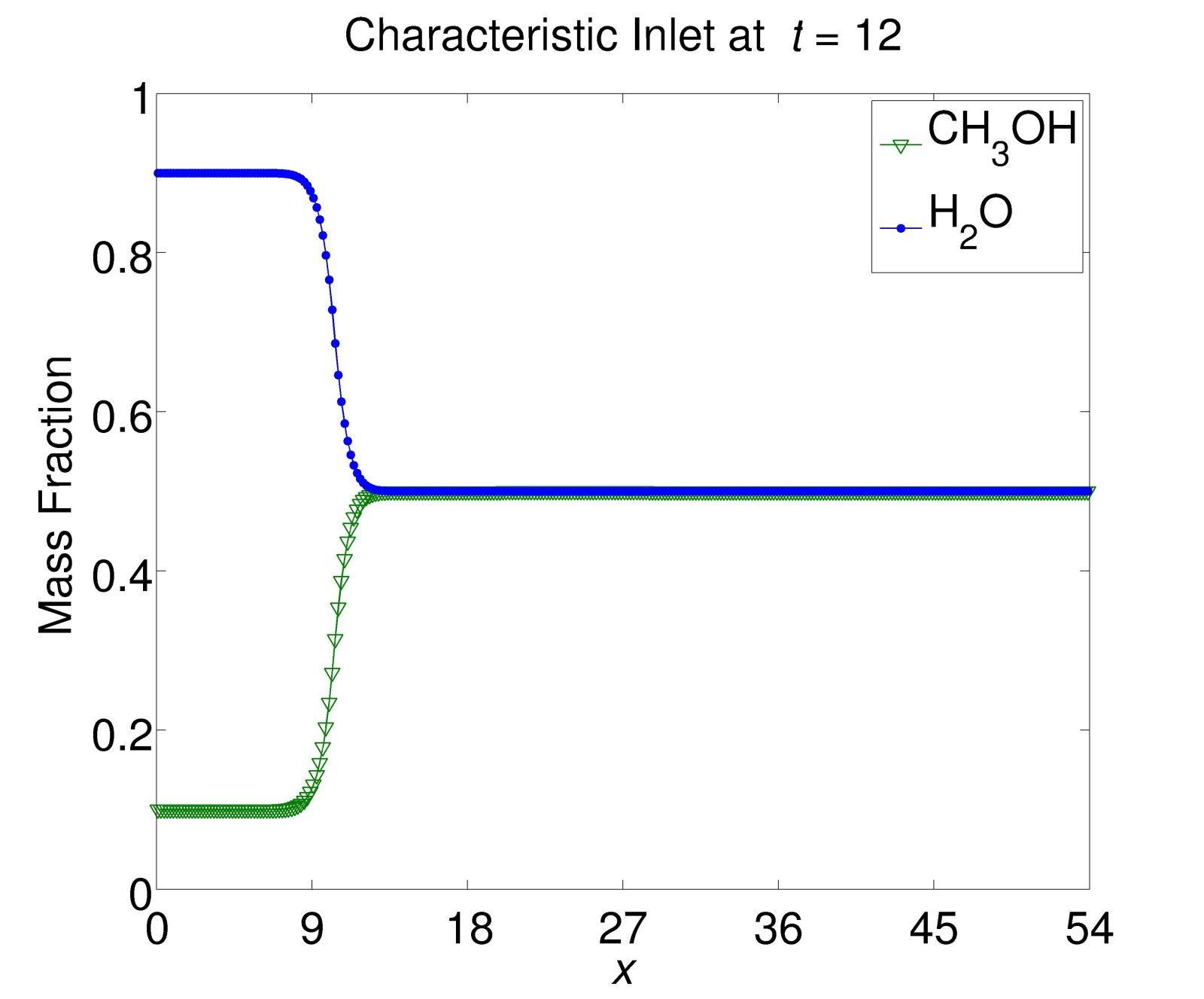}\hspace{-10pt}\includegraphics[width=6.2cm]{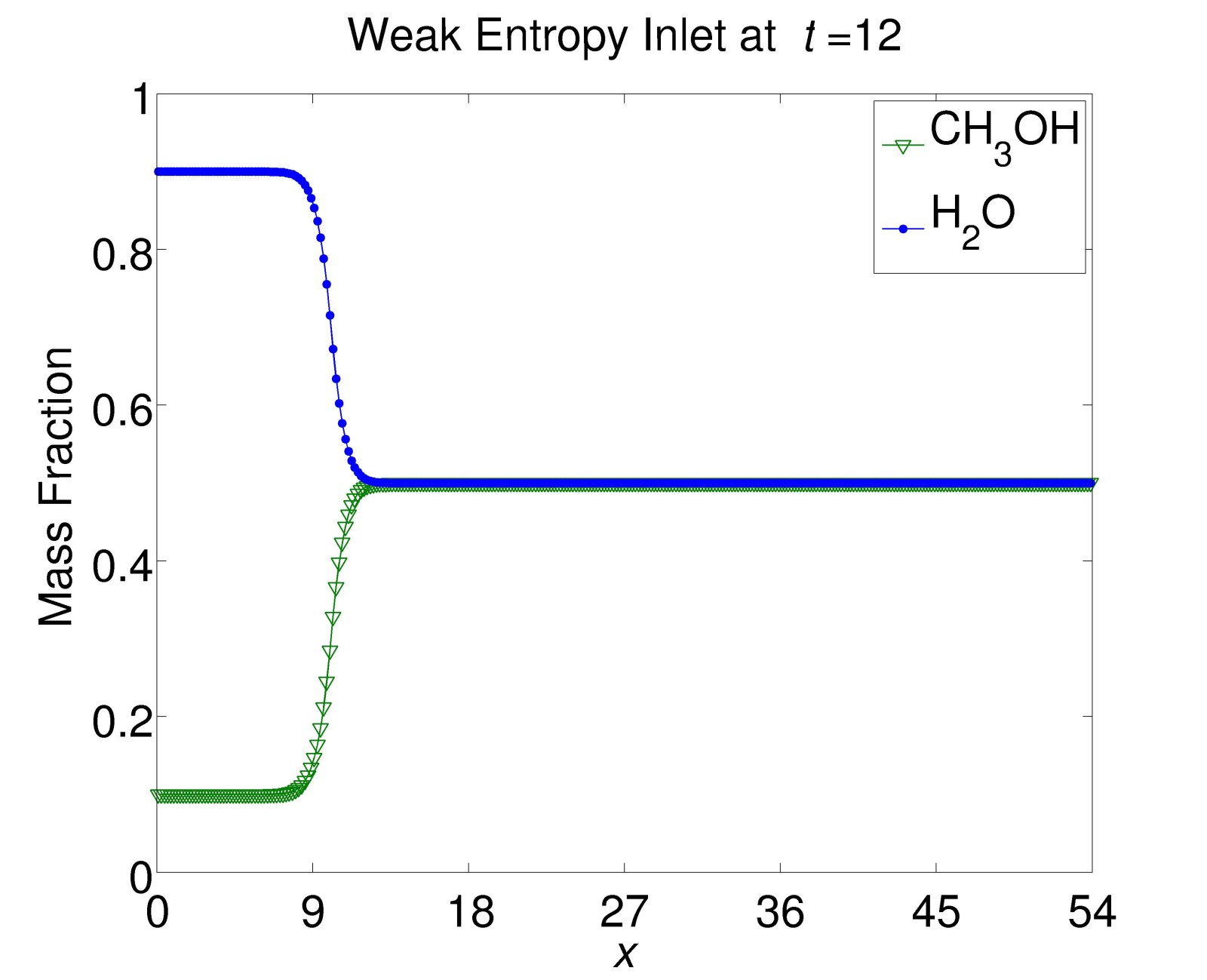}
\caption{The left plot shows miscible species at $t=12$ given the characteristic chemical inlet conditions from (\ref{cheminsys}) with $\mathscr{C}=0.9$ and on the boundary $a = 0$, with the first order transmissive conditions on $b=x_{ne}$ (see Figure \ref{fig:scheme}).  The right plot shows the same solution using the weak entropy formulation.  Here we have a miscible solution of methanol and water at $\vartheta=500$K and initial $\rho_{0} =5$, $u_{0}=0$, and $\mu_{1,0}=\mu_{2,0}=0.5$.}
\label{fig:mudift}  
\end{figure}

Now as in \textsection{4} we easily recover the form \begin{equation}\boldsymbol{U}_{t} + \boldsymbol{\Gamma}\boldsymbol{U}_{x} = (\mathscr{K}\boldsymbol{U}_{x})_{x},\end{equation} which integrates to (\ref{weak}).  Again we solve for our system in a form equivalent to (\ref{aprox}).   We employ the local Lax-Friedrich's inviscid flux $\boldsymbol{\Phi}_{lLF}$, the Bassi-Rebay numerical flux $\hat{\boldsymbol{U}}_{BR}$, the usual viscous flux $\hat{\mathscr{G}}_{b}$, and the van Leer slope limiter.


\begin{figure}[t!]
\centering
\includegraphics[width=9cm]{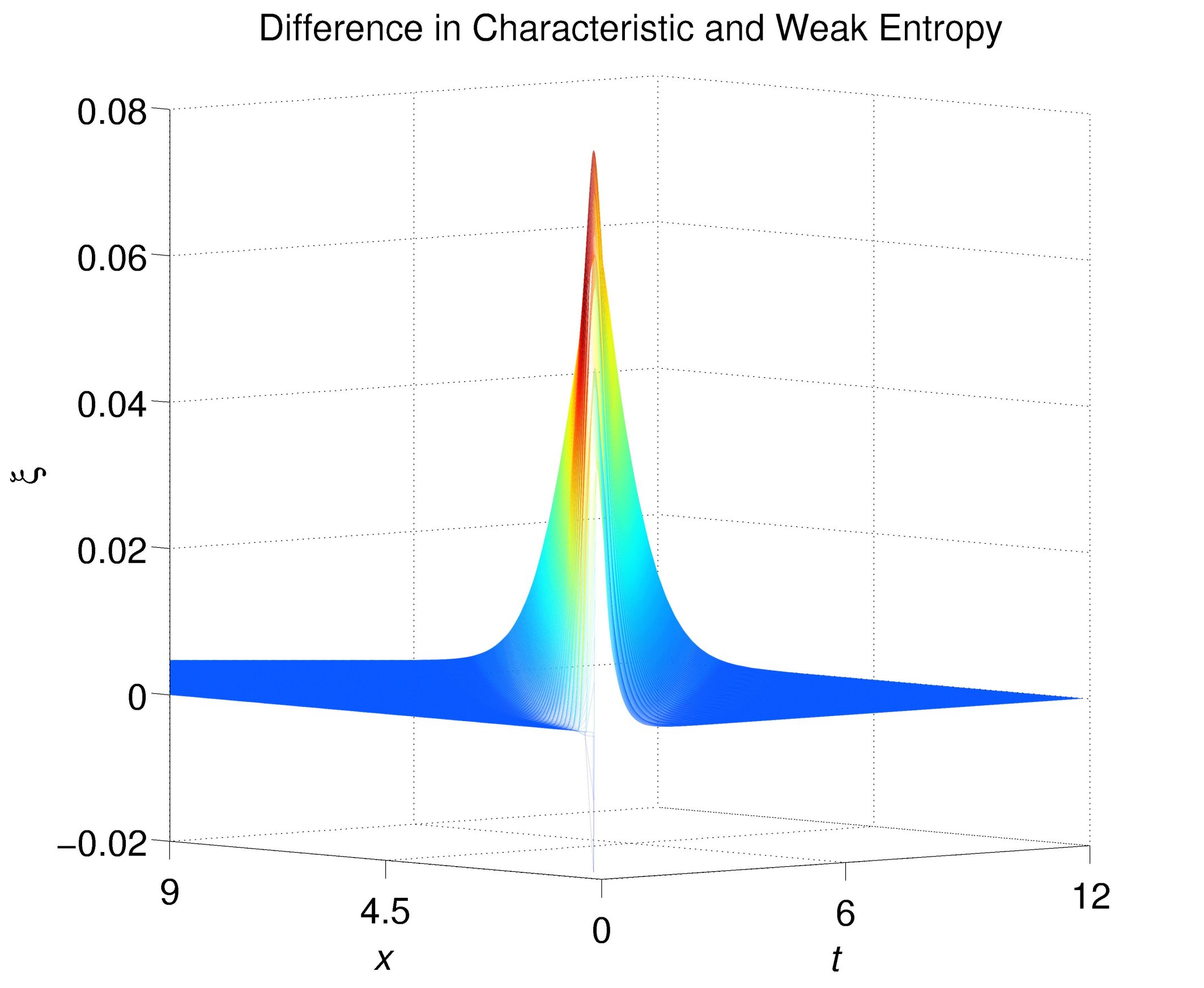}
\caption{Here we show the difference between the first chemical constituent of the weak entropy $\mu_{1,w}$ and characteristic $\mu_{1,c}$ solutions shown in Figure \ref{fig:mudift}, where $\xi = \mu_{1,w}-\mu_{1,c}$.  In this figure, for emphasis, we show only the reduced spatial interval $(0,9)$.}
\label{fig:mudif}  
\end{figure}

All that remains is determining the boundary states $\boldsymbol{U}_{h}^{n}|_{\partial_{\Omega}}$.  We begin by considering the case of characteristic boundary conditions, and assume that at the boundary $x=a$ we have a subsonic inlet $u\cdot\boldsymbol{n} < 0$.  In our determination of characteristic boundary conditions, we linearize about the state $\boldsymbol{U}_{h}^{n}|_{K_{ij}}$ to arrive at an expression for the boundary state $\boldsymbol{U}_{h}^{n}|_{K_{ji}}$ at timestep $t^{n}$.  Then, from Table \ref{table:odesys}, we see that $\beta_{1}^{n}$ is fixed by \begin{equation} \label{eq:beta_1} \beta_{1}^{n} = \rho^{n}(a^+)/2, \ \ \mathrm{where} \ \ \rho^{n}(a^+) = \lim_{x \rightarrow 0^+} \rho^{n}(a+x). \end{equation}  Now, suppose we want a chemical inlet such that the first chemical constituent $\mu_{1}$ is characterized by an influx condition $\mu_{1}^{n}(a^-)=\mathscr{C}$ where similarly, \[ \mu^{n}(a^-) = \lim_{x \rightarrow 0^+} \mu^{n}(a-x). \]  In order to maintain the consistency of our system, we additionally need that $\mu_{2}^{n}(a^-)=1-\mathscr{C}$ and $\rho^{n}(a^-)=\epsilon>0$.   For $\ell=2$, we solve (\ref{alpbet}) with the constraint in (\ref{eq:beta_1}) to obtain:
\begin{equation}
\begin{aligned}
\label{cheminsys}
\beta_{2}^{n}=-\rho^{n}(a^+)/2 - \beta_{3}^{n} &+\epsilon, \quad\quad \beta_{3}^{n} = \beta_{4}^{n}\xi^{n}(a^+), \\ \beta_{4}^{n} = \epsilon(\mu_{1}^{n}(a^+)-\mathscr{C})/c_{n}^{2}(a^+), \ &\textup{  and  } \ (\rho u)^{n}(a^-) = \beta_{1}^{n}+\beta_{2}^{n}+\beta_{3}^{n},
\end{aligned}
\end{equation} where $\xi^{n}(a^+) = \partial_{\rho_{1}}p^{n}(a^+) - \partial_{\rho_{2}}p^{n}(a^+),$ $c_{n}(a^+)$ denotes the speed of sound at timestep $t^{n}$ on the boundary as defined in the appendix.  Finally, at the other boundary point $x=b$ we set a transmissive characteristic boundary condition.

\begin{figure}[t!]
\centering
\includegraphics[width=9cm]{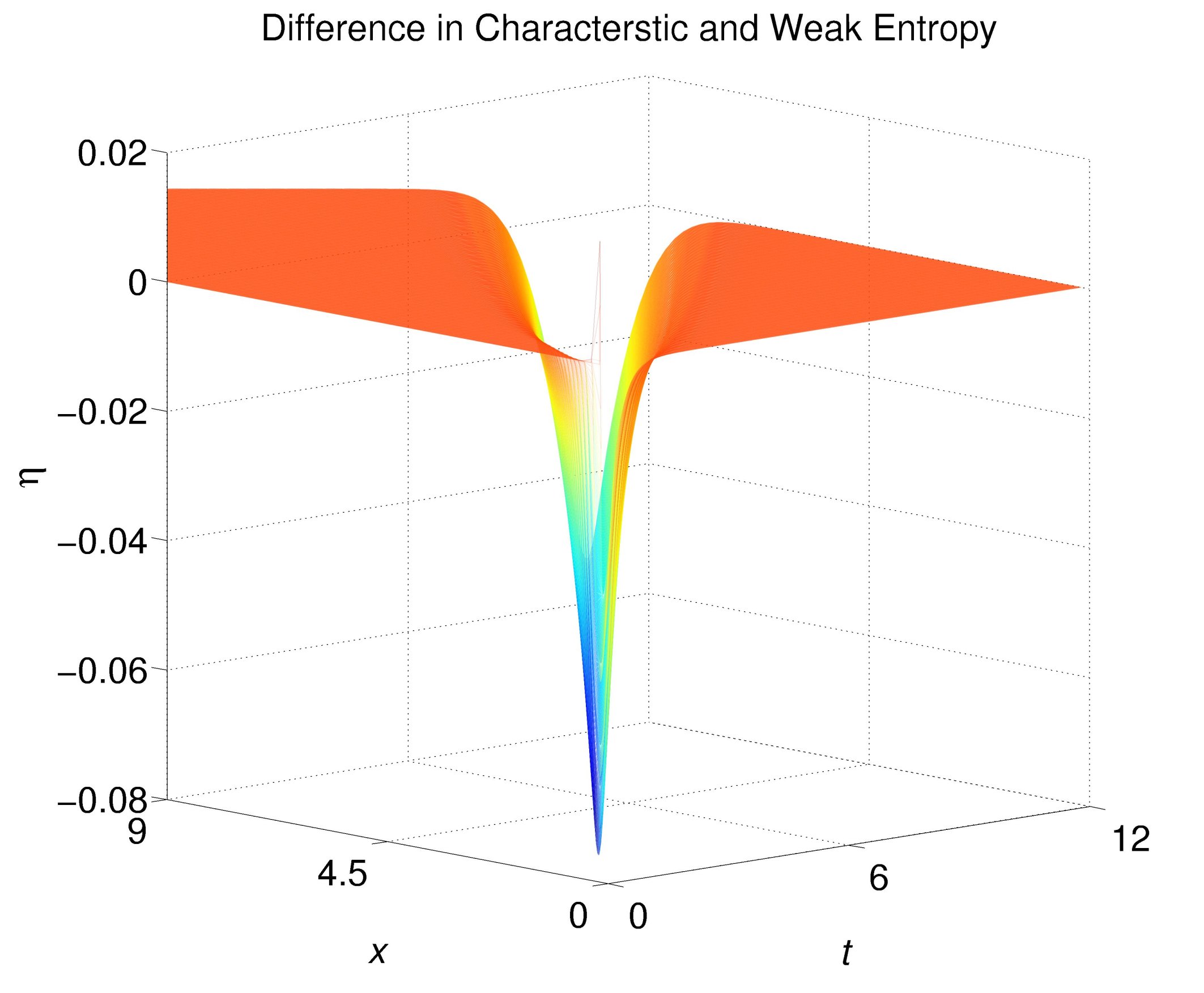}
\caption{Here we have the complementary difference between species two of the weak entropy $\mu_{2,w}$ and characteristic $\mu_{2,c}$ solutions, where $\eta = \mu_{2,w}-\mu_{2,c}$}
\label{fig:mudif2}  
\end{figure}

The behavior of such a ``chemical inlet'' is shown in Figure \ref{fig:mudift} where we have utilized a mesh size of $h=0.54$ and a timestep of $h/30$.  Here we set $\epsilon=\rho_{0}(a^+)$.   By comparison the weak entropy solutions discussed in \textsection{3} to (\ref{mass2})-(\ref{specie2}) are also well-posed for an arbitrary collection of $L^{\infty}((0,T)\times\partial\Omega)$ boundary data.  So, in contrast to decomposing the solution into its characteristic directions, we may simply assign $\mu_{1}^{n}(a^-)=\mathscr{C}$, $\mu_{2}^{n}(a^-)=1-\mathscr{C}$, $\rho^{n}(a^-)=\epsilon$ and the lag velocity condition $u^{n}(a^-)=u^{n-1}(a^+)$ for every timestep to obtain the weak entropy solution.  

Comparing the behavior of the weak entropy solution of the mass fraction of the first constituent $\mu_{1,w}$ in Figure \ref{fig:mudift} to the characteristic solution of the mass fraction of the first constituent $\mu_{1,c}$ yields Figure \ref{fig:mudif}.  Notice that the two boundary solutions do not demonstrate the same numerical behavior.  In particular, the weak entropy $\mu_{1}$ grows more rapidly at the boundary; while the dynamically coupled characteristic solution adapts to the influx of specie/density by producing a velocity outflow, which effectively reduces the ``chemical influx'' as a function of time.  

In practice it is often physically meaningful to ascribe more boundary data than the free characteristic directions associated to the free $\beta$'s can consistently control.  For example a closely related case to the chemical inlet example given above, is the subsonic outlet $u_{b}\cdot \boldsymbol{n} > 0$.  In cases such as these, where only one characteristic direction is free, weak entropy solutions are essential in order to even characterize such mixing at the boundary interface (such a case emerges of particular interest, for example, when interspecies diffusion occurs in the mass transport as shown in \textsection{8}).  Heuristically we may say that characteristic solutions demonstrate a relatively weaker forcing on $\partial\Omega$ but are more restictive in terms of degrees of freedom, while the weak entropy boundary solutions display a greater flexibility of representation by way of establishing stronger forcing on $\partial\Omega$.

\section{\texorpdfstring{\protect\centering $\S 6$ Example: $k$-th order in time $\ell$-fluid}{\S 6 Example: k-th order in time l-fluid}}

We wish to generalize the example in \textsection{5} to $\ell$-fluid components and a $k$-th order in time Runge-Kutta time discretization.  Let us start with an $\ell=5$ system, which then can be easily generalized.  Consider
\begin{align}
\label{mass5}&\partial_{t}\rho+\partial_{x}(\rho u)=0,\\
\label{momentum5}&\partial_{t}(\rho u)+\partial_{x}(\rho u^{2})+\partial_{x}p -\partial_{x}(\nu\partial_{x}u)=0, \\
\label{specie5}&\partial_{t}(\rho\mu_{i})+\partial_{x}(\rho u\mu_{i})=0,
\end{align}
with initial conditions, 
\[
\rho_{|t=0}=\rho_{0}>0,\qquad \rho u_{|t=0}=m_{0}, \qquad\mathrm{and}\qquad (\rho\mu_{i})_{|t=0}=\rho_{i,0},
\]
given the pressure
\begin{equation}
\label{pressure5}
p= \rho_{1}^{\gamma_{1}}+ \rho_{2}^{\gamma_{2}} +  \rho_{3}^{\gamma_{3}} +  \rho_{4}^{\gamma_{4}} +  \rho_{5}^{\gamma_{5}},
\end{equation}
and viscosity
\begin{equation}
\label{viscosity5}
\nu= \psi'(\rho_{1}\partial_{\rho_{1}}p +  \rho_{2}\partial_{\rho_{2}}p+  \rho_{3}\partial_{\rho_{3}}p+  \rho_{4}\partial_{\rho_{4}}p +  \rho_{5}\partial_{\rho_{5}}p).
\end{equation}

\begin{figure}[t!]
\centering
\includegraphics[width=6.2cm]{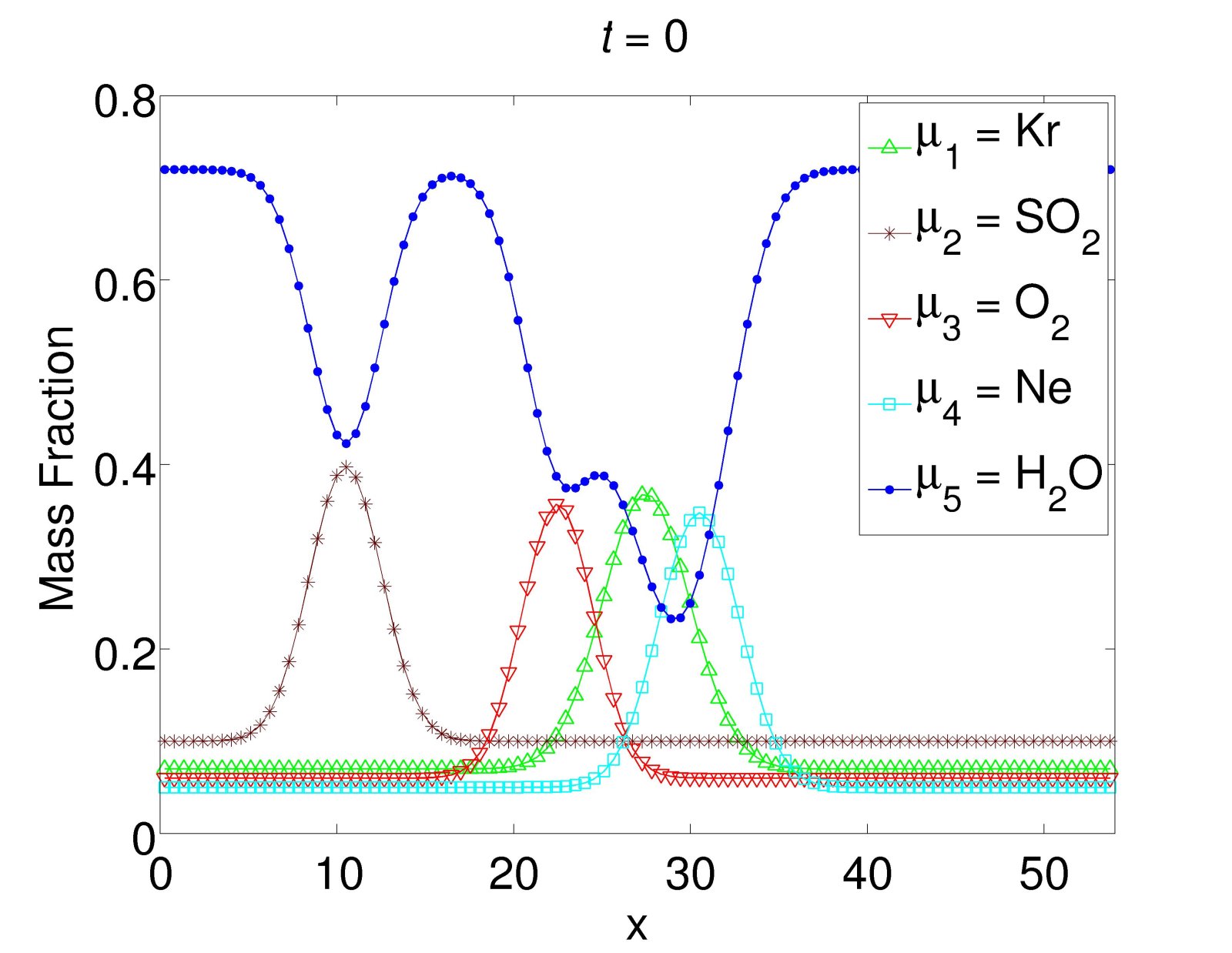}\includegraphics[width=6.2cm]{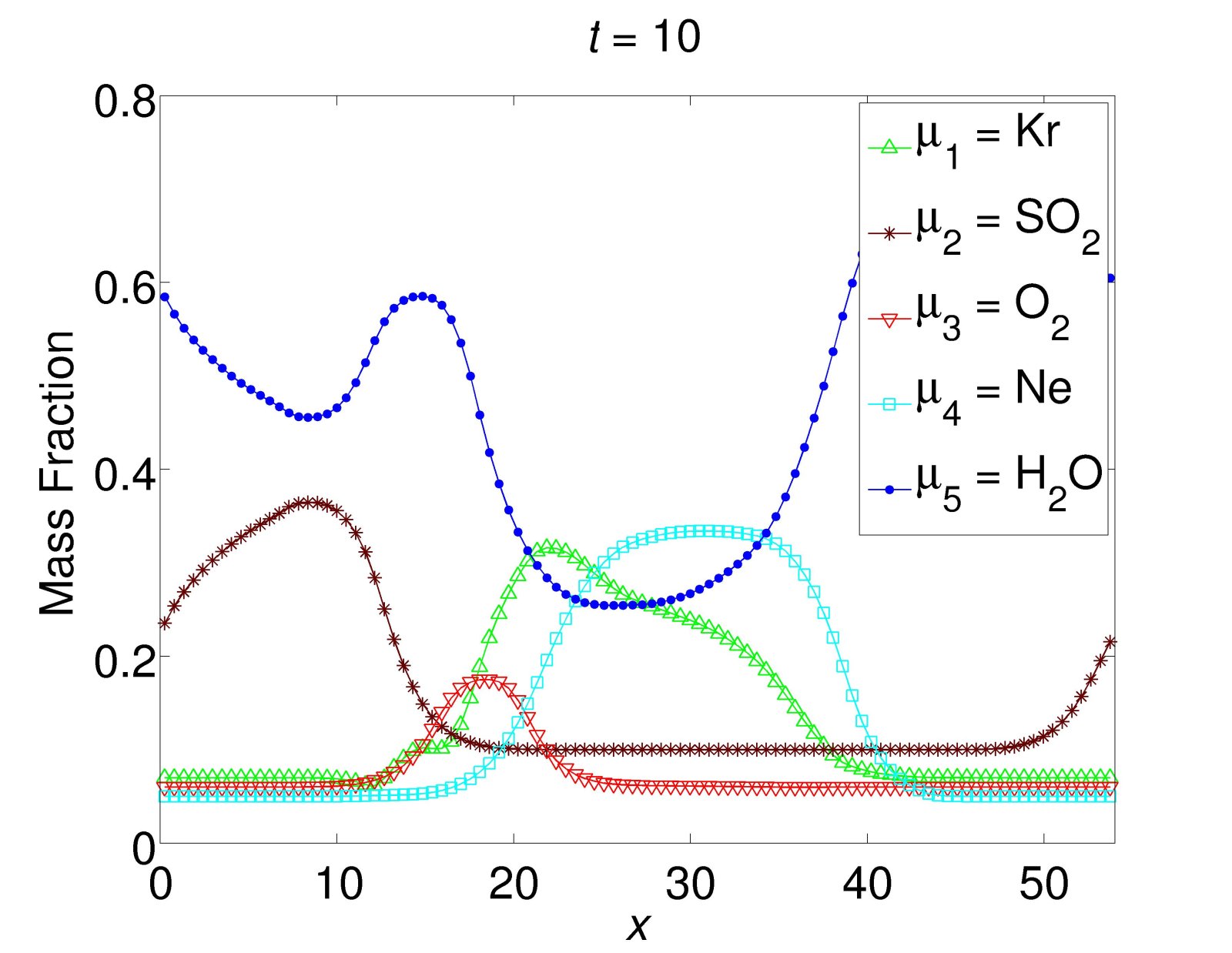}
\caption{Here we show the first and last timesteps of the mass fractions at $\vartheta= 293$K using periodic boundary conditions with Runge-Kutta order $k=2$. Initial conditions set $\rho= 5 + 20e^{-(x - 10)^{2} / 8} + 20e^{-(x - 30)^{2}/8}$ and $u =\sin(6\pi x/x_{ne})$, with $\mu_{1}=0.07 + 0.3e^{-(x - 27.5)^{2}/ 12}$, $\mu_{2}=0.1 + 0.3e^{-(x - 10.5)^{2}/8}$, $\mu_{3}=0.06 + 0.3 e^{-(x - 22.5)^{2}/8}$, $\mu_{4}=0.05 + 0.3e^{-(x - 30.5)^{2}/10}$ and solvent $\mu_{5}=1-\sum_{n=1}^{4}\mu_{i}$.}
\label{fig:n5k3}  
\end{figure}

We take the three vectors $\boldsymbol{U}= (\rho, \rho u,\rho_{1},\rho_{2},\rho_{3},\rho_{4},\rho_{5})^{T},$ $\boldsymbol{f}(\boldsymbol{U})=(\rho u, \rho u^{2}+p, \rho_{1} u,\rho_{2} u, \rho_{3} u, \rho_{4} u, \rho_{5}u)^{T},$ and \mbox{$\boldsymbol{g}(\boldsymbol{U},\boldsymbol{U}_{x}) = (0,\nu u_{x}, 0, 0, 0, 0, 0, 0)^{T}$}, such that again we arrive with \begin{equation}\label{matform3}\boldsymbol{U}_{t}+\boldsymbol{\Gamma}\boldsymbol{U}_{x}-(\mathscr{K}\boldsymbol{U}_{x})_{x}=0,\end{equation} which is easily approximated by the numerical scheme given in (\ref{aprox}).  

We generalize to higher order time discretization.  That is, let us rewrite (\ref{aprox}) as a system of ordinary differential equations,
\[
\frac{d}{dt}\boldsymbol{U}_{h}=L_{h}(\boldsymbol{U}_{h}).
\]
\begin{figure}[t!]
\centering
\includegraphics[width=9cm]{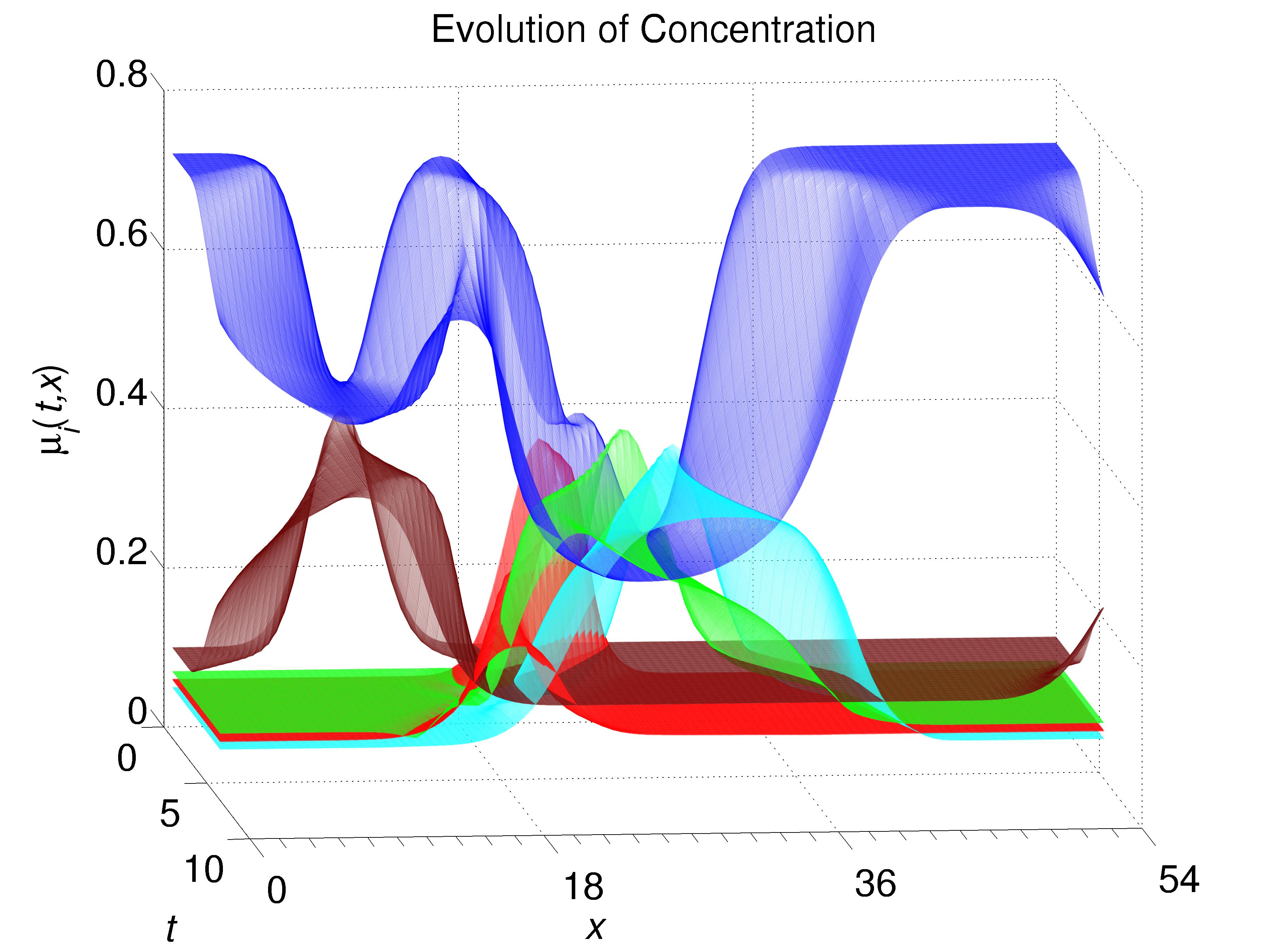}
\caption{Here we show the time evolution over the entire solution space of the same problem from Figure \ref{fig:n5k3}.}
\label{fig:n5k32}  
\end{figure}
We can solve this system using an explicit Runge-Kutta method. Specifically, we use the strong-stability preserving Runge-Kutta methods presented in Ref.~\cite{Cockburn1}. 
This method follows for any $\ell$-fluid of the form (\ref{mass})-(\ref{viscosity}) of Runge-Kutta order $k$.  


The behavior of this system is shown in Figures \ref{fig:n5k3} and \ref{fig:n5k32}, where we have set the simple periodic boundary condition,
\[
\boldsymbol{U}_{h}^{n}(a^{+},t)=\boldsymbol{U}_{h}^{n}(b^{+},t), \hspace{10pt} \boldsymbol{U}_{h}^{n}(a^{-},t)=\boldsymbol{U}_{h}^{n}(b^{-},t) .
\]
The numerical solution shown was obtained using a mesh size of $h=0.36$, a timestep of $h/30$, the local Lax-Friedrich's flux, van Leer's slope limiter, and the Runge-Kutta method presented in Ref.~\cite{Cockburn1} with $k=2$. It is worth noting that the composition of this mixture does not tend towards homogeneous equilibrium, since there is both no interspecies diffusion (see \textsection{8}) and the species are not ``chemically miscible'' (in that they do not mix in all proportions).  Nevertheless there is significant mixing from the state of the initial conditions, and it can be seen that the fluid is \emph{more} homogenized, relatively speaking, at time $t=10$ that it was in the initial state.  Most importantly, this scheme now immediately extends to an arbitrary $\ell$-fluid.   


\section{\texorpdfstring{\protect\centering $\S 7$ Energy consistency of scheme}{\S 7 Energy consistency of scheme}}

In Ref.~\cite{MV} it is shown that any solution for which the Theorem holds should satisfy two closely related entropy inequalities.  The first, a classical integral inequality taking the form
\begin{equation}
\begin{aligned}
\label{energy1}
\frac{1}{2}\frac{d}{dt}\int_{\mathbb{R}}\big\{\rho u^{2}+2\mathscr{E}\big\}dx+\int_{\mathbb{R}}\nu|u_{x}|^{2}dx\leq 0,
\end{aligned}
\end{equation}
and the second owing to Bresch and Desjardins (see Ref.~\cite{BD2,BD3}), as 
\begin{equation}
\label{energy2}
\frac{1}{2}\frac{d}{dt}\int_{\mathbb{R}}\big\{\rho|u + \rho^{-1}\psi_{x}|^{2} + 2\mathscr{E}\big\}dx+\int_{\mathbb{R}}\rho^{-1}\psi'|p_{x}|^{2} dx \leq 0,
\end{equation}
where the internal energy $\mathscr{E}=\mathscr{E}(\rho_{1},\ldots,\rho_{n})$ is specified as: 
\[
\mathscr{E} = \sum_{i=1}^{\ell}\frac{\rho_{i}^{\gamma_{i}}}{\gamma_{i}-1}.
\]

Entropy consistent numerical schemes are often formulated in the literature in order to explicitly enforce entropy inequalities such as (\ref{energy1}) and (\ref{energy2}) over all of $\mathcal{Q}_{T}$ (viz. Ref.~\cite{Barth2,Hughes1}).  For example enforcing (\ref{energy1}) may be done by utilizing a change of variables of the conservation variable form of the state vector $\boldsymbol{U}$, into the so-called entropy variable form $\boldsymbol{W}$, which is achieved by writing the entropy functional $\mathscr{H}=\rho u^{2}/2 + \mathscr{E}$ and then setting the state vector as the partial with respect to the conservation variables $\boldsymbol{W}=\mathscr{H}_{\boldsymbol{U}}$.  The difficulty of implementation of these energy schemes, which are inherently implicit methods, underscores the importance of conserving energy consistency of the solution, and further serves as motivation for testing how our explicit scheme behaves with respect to (\ref{energy1}) and (\ref{energy2}).  

\begin{figure}[t]
\centering
\includegraphics[width=9cm]{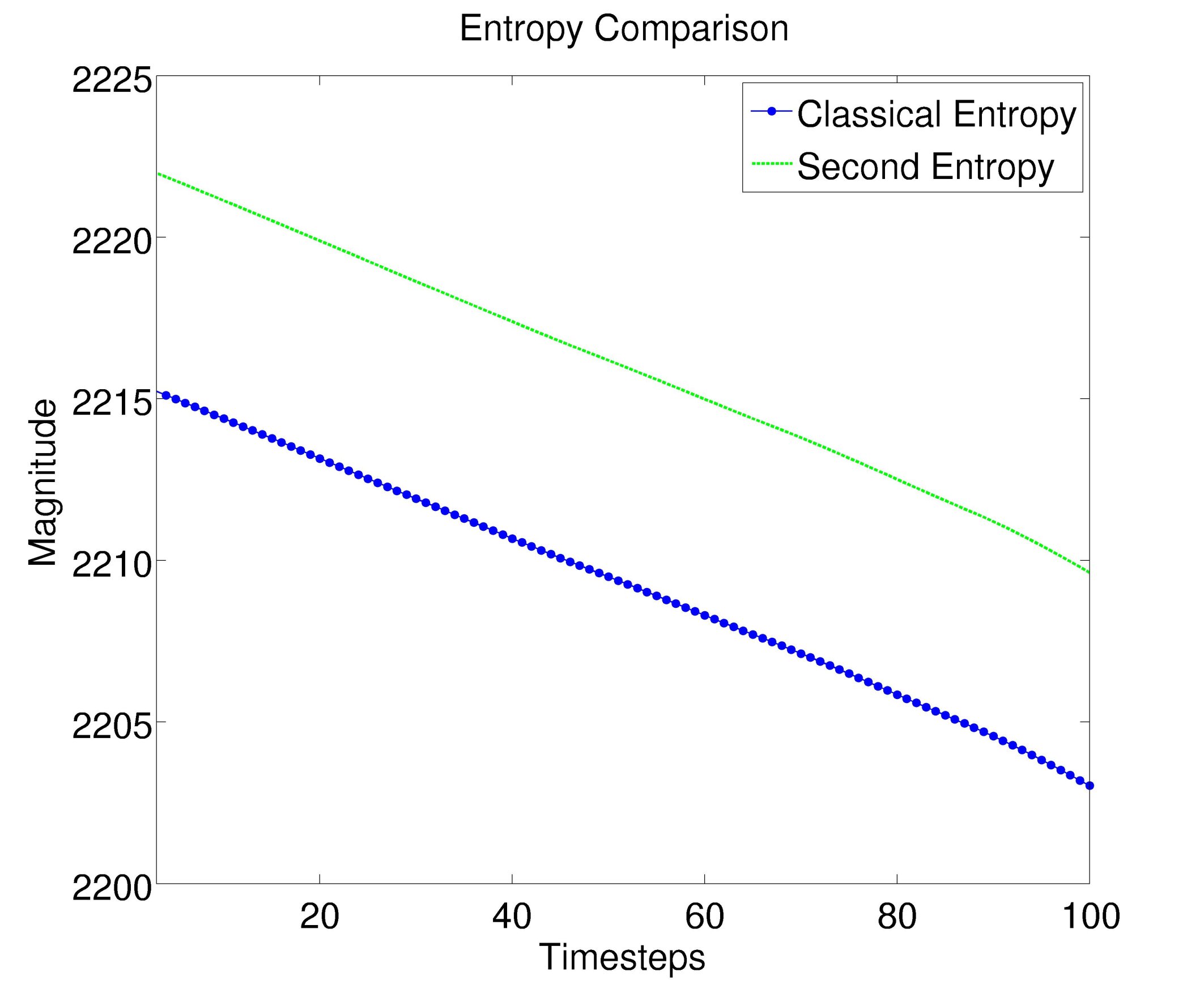}
\caption{Here we plot the integral forms $\mathscr{S}_{T}$ and $\tilde{\mathscr{S}}_{T}$ for $C=1$ and $\alpha = 0.9$, where $\int_{\Omega}\mathscr{H}_{0}dx$ and $\int_{\Omega}\tilde{\mathscr{H}}_{0}dx$ are represented by the first timestep.  The spatial mesh is chosen with $ne=100$ with $\Delta t = 0.01$.}
\label{fig:CompEntropy}  
\end{figure}

Here we inspect the entropy consistency of our scheme with respect to (\ref{energy1}) and (\ref{energy2}) using the $\ell=5$ fluid with periodic boundary data as shown in \textsection{6}.  From the numerical perspective, we expect our solution (\ref{aprox}) to obey entropy consistency up to a restriction of the CFL stability condition, which for inviscid flows scale as $\tilde{C}_{1}h/\Spec_{r}\Gamma\geq \Delta t$ and for the complementary viscous flows like $\tilde{C}_{2}h^{2}/\max(\nu,1)\geq\Delta t$, where the CFL constants are characterized by $\tilde{C}_{1},\tilde{C}_{2}\in (0,1)$. We note that we do not expect energy consistency for an arbitrary choice of boundary data.

\begin{figure}[t!]
\centering
\includegraphics[width=9cm]{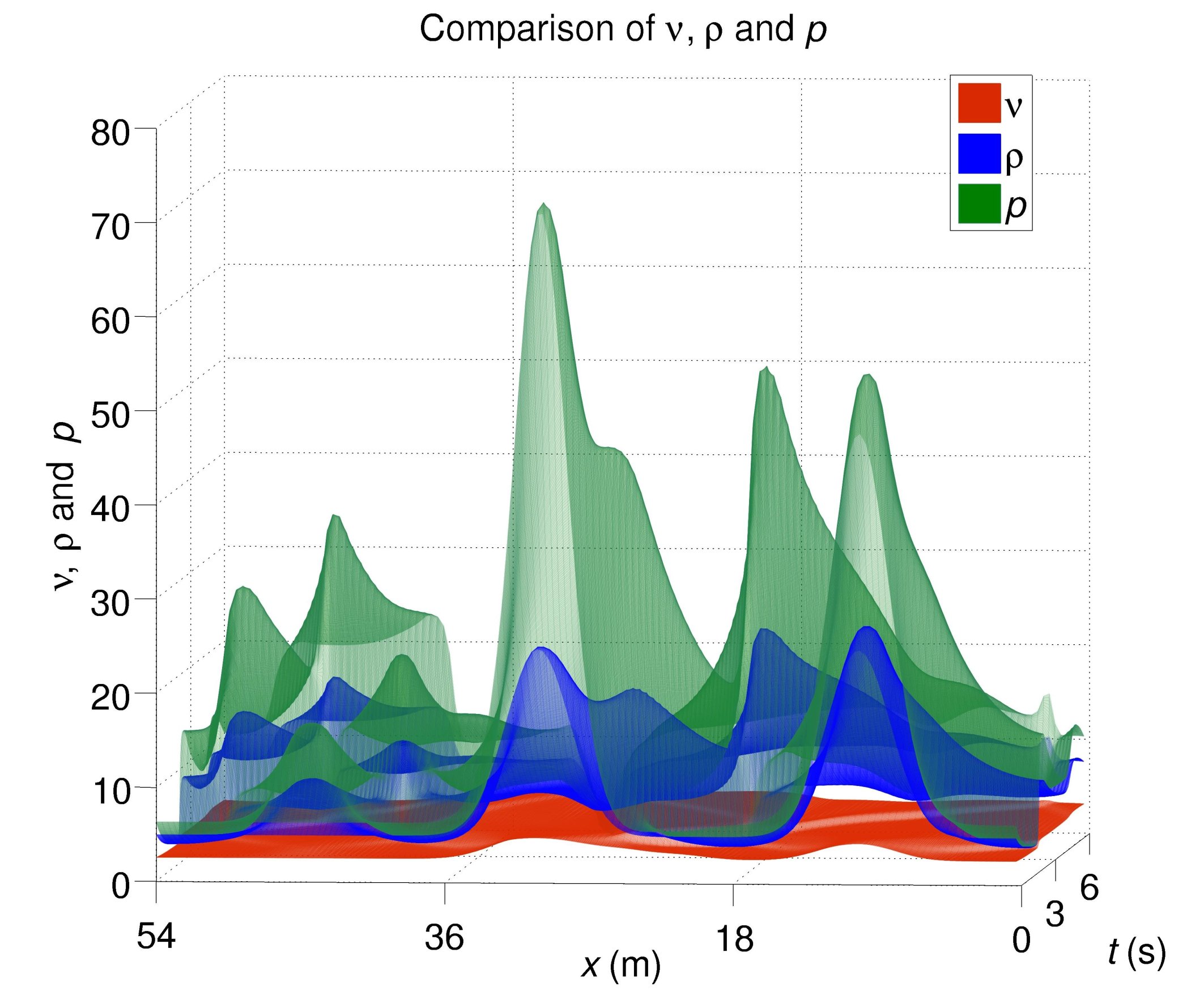}
\caption{Here we compare the viscosity $\nu$, density $\rho$ and pressure $p$ of the periodic 5-fluid from \textsection{6} with $\alpha=0.9$, $C=0.5$, $150$ meshpoints and $\Delta t =.006$.  }
\label{fig:ViscRhoPressure}  
\end{figure}

To examine whether the two inequalities (\ref{energy1}) and (\ref{energy2}) are satisfied, we first define $\tilde{\mathscr{H}}=\frac{1}{2}\rho|u + \rho^{-1}\partial_{x}\psi|^{2} + \mathscr{E}$, and check that the spacetime integrated functionals for our numerical solution satisfy: \begin{equation}\label{energy12}\mathscr{S}_{T} = \int_{\Omega}\mathscr{H}_{T}dx+\int_{0}^{T}\int_{\Omega}\nu|u_{x}|^{2}dx dt\leq \int_{\Omega}\mathscr{H}_{0}dx,\end{equation} and \begin{equation}\label{energy22}\tilde{\mathscr{S}}_{T}= \int_{\Omega}\tilde{\mathscr{H}}_{T}dx+\int_{0}^{T}\int_{\Omega}\rho^{-1}\psi'|p_{x}|^{2} dx dt\leq \int_{\Omega}\tilde{\mathscr{H}}_{0}dx.\end{equation}

We show the results of this calculation for an arbitrarily chosen set of parameters in Figure \ref{fig:CompEntropy}.  As is clear from the graph, both (\ref{energy12}) and (\ref{energy22}) are satisfied.  In fact we have confirmed that (\ref{aprox}) satisfies (\ref{energy12}) and (\ref{energy22}) whenever the CFL condition is satisfied, up to the choice of a constant.  It is interesting to note that both of these inequalities are satisfied for an arbitrary choice of $\alpha$ and $C$ in the numerical setting.  This confirms that the mathematical result from Ref.~\cite{MV} is substantially more restrictive than the numerical one.


As a side remark, the functional behavior of the viscosity is a relatively unique property of our system (\ref{mass})-(\ref{specie}), which is to say that commonly compressible Navier-Stokes systems utilize constant viscosity coefficients (eg. see Ref.~\cite{FFS} chapter 4 and Figure \ref{fig:ViscRhoPressure}) and thus the energy consistency and the CFL condition is not dynamically coupled to the solution components.  However, for our system, since the viscosity is a function of time, the CFL condition must update to reflect the local viscosity magnitude at each timestep.  

\section{\texorpdfstring{\protect\centering $\S 8$ Fick's diffusion with acoustic BCs}{\S 8 Fick's diffusion with acoustic BCs}}

\begin{figure}[b!]
\centering
\includegraphics[width=10cm]{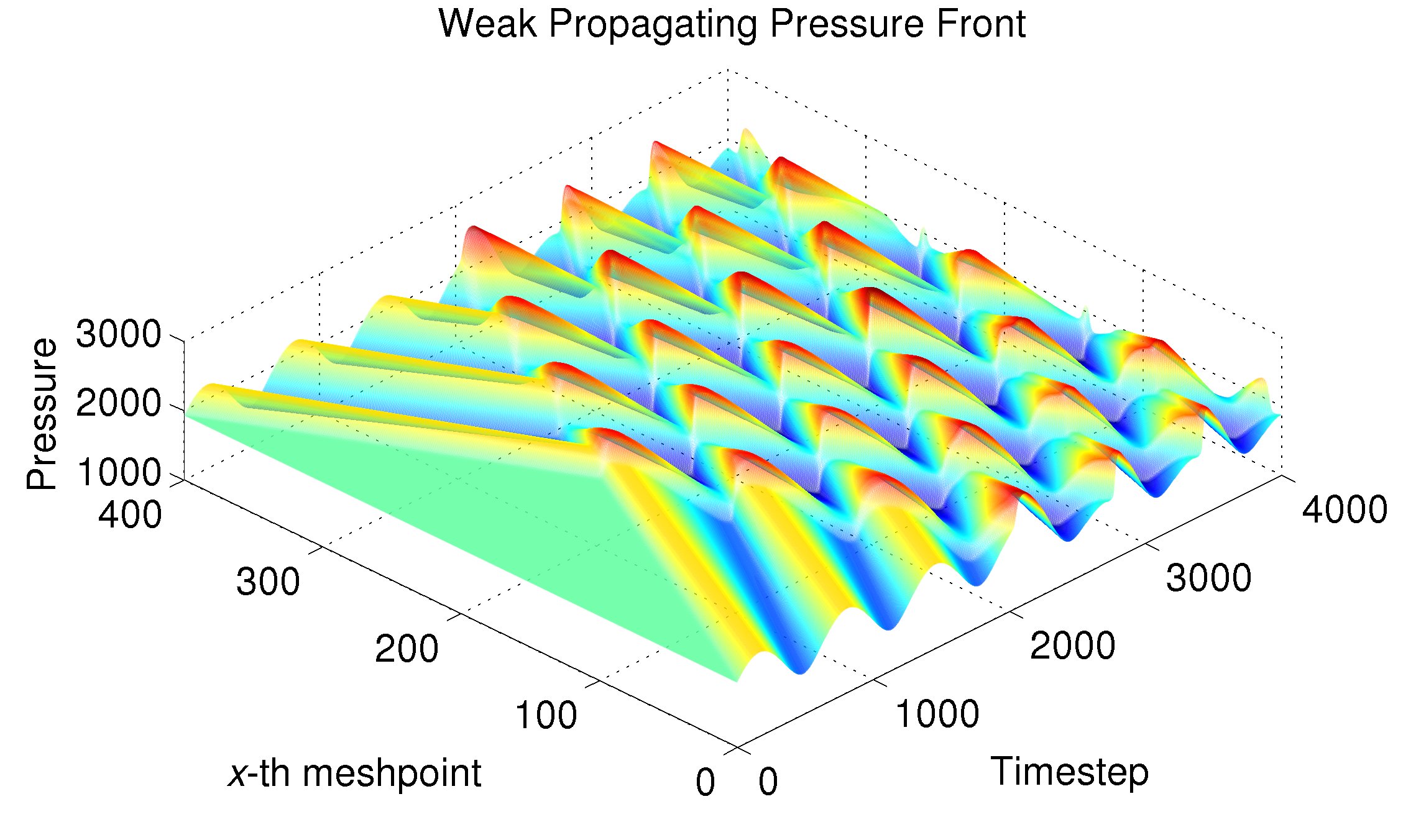}
\caption{A weak entropy solution to an oscillating pressure front propagating through a $5$-component low density ($\sim 100$ molecules per cm) gas at $\vartheta = 20$K.  The chemical constituents are comprised of species found in dark interstellar molecular clouds, where representative fractional abundances are adopted and the solution space is appropriately scaled; with corresponding initial conditions: $\mathrm{H}_{2}\sim 80\%$, $\mathrm{He}\sim 19.9\%$, and trace $\mathrm{CO}$, $\mathrm{H}$ (atomic hydrogen), and $\mathrm{HC}_{3}\mathrm{N}$ (cyanoacetylene).}
\label{fig:interstellar}  
\end{figure}

Although Theorem 2.1 only applies to systems of the form (\ref{mass})-(\ref{specie}), the particular numerical scheme outlined in \textsection{2} can be easily extended to more complicated systems; and indeed can be extended with similar numerical behaviors.  As an example let us consider the $5$-fluid,
\begin{align}
\label{massf}&\partial_{t}\rho+\partial_{x}(\rho u)=0,\\
\label{momentumf}&\partial_{t}(\rho u)+\partial_{x}(\rho u^{2})+\partial_{x}p -\partial_{x}(\nu\partial_{x}u)=0, \\
\label{specief}&\partial_{t}(\rho\mu_{i})+\partial_{x}(\rho u\mu_{i}) - \partial_{x}(\rho\mathscr{D}_{i}\partial_{x}\mu_{i}) = 0,
\end{align}
with initial conditions: 
\[
\rho_{|t=0}=\rho_{0}>0,\qquad \rho u_{|t=0}=m_{0}, \qquad\mathrm{and}\qquad (\rho\mu_{i})_{|t=0}=\rho_{i,0},
\]
given (\ref{pressure5}), (\ref{viscosity5}) and $\mathscr{D}_{i}$ the diffusivity constants of each respective species.  Here the system is equivalent to that in \textsection{6}, except we have added the Fick's diffusion law term to the advection equation in $\mu$.  Thus the state vector and inviscid flux remain unchanged, while the vector $\boldsymbol{g}$ becomes \begin{equation}\label{fic}
\boldsymbol{g}(\boldsymbol{U},\boldsymbol{U}_{x}) = (0,\nu u_{x}, \rho\mathscr{D}_{1}\mu_{x}, \ldots , \rho\mathscr{D}_{n}\mu_{\ell})^{T},
\end{equation}
such that the corresponding viscous flux matrix yields $\mathscr{K}=\partial_{\boldsymbol{U}_{x}}\boldsymbol{g}$.
   


We set an acoustic inlet condition, which is equivalent to identifying the sound pressure on $\partial\Omega$.  We suppose that the pressure on the boundary is a classical time-harmonic solution to the acoustic wave equation, namely, $p_{b}= p_{0} + A_{0}\sin(\omega t)$ for a driving amplitude $A_{0}$, an ambient reference pressure $p_{0} = \sum_{i}^{\ell}(\rho_{0}\mu_{i,0})^{\gamma_{i}}$, and an angular frequency $\omega$.  

Here we have solved (\ref{massf})-(\ref{specief}) using a formulation which is meant to weakly mimic some of the conditions found in interstellar nurseries, or interstellar molecular clouds.  The solution is shown in Figure \ref{fig:interstellar}, where it is notable that the traveling sound field $p_{b}$ dynamically responds to the changing speed of sound $c$ throughout the medium -- which scales like the root of the local change in pressure up to the local species concentration.  The initial conditions and the diffusivities were estimated with the help of Ref.~\cite{RH,MdRW}.

\begin{figure}[t!]
\centering
\includegraphics[width=10cm]{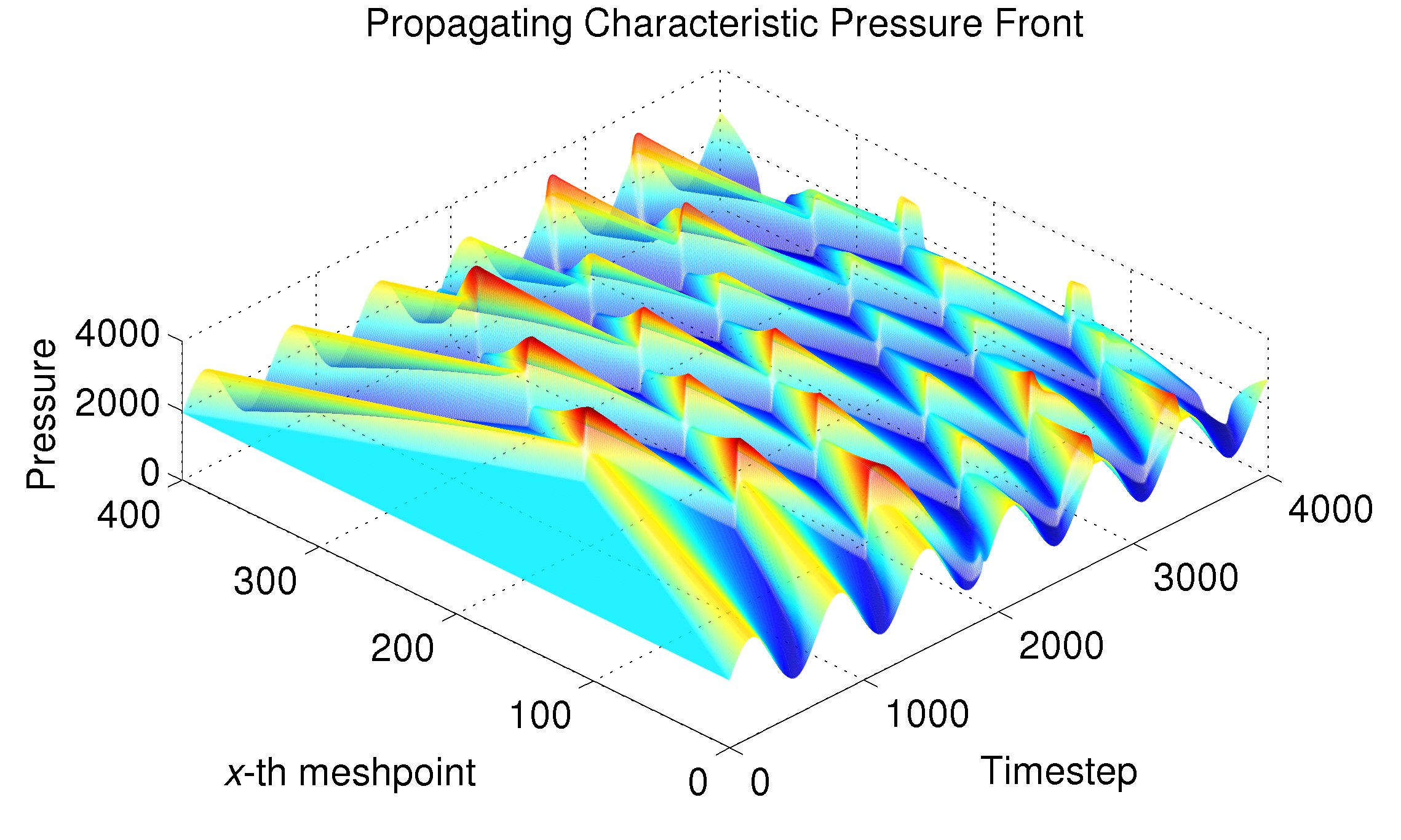}
\caption{A characteristic solution to the same oscillating pressure front presented in \ref{fig:interstellar}.}
\label{fig:charinterstellar}  
\end{figure}

For the case of weak entropy conditions, we set the lag velocity condition $u_{b}^{n}|_{\mathcal{K}_{ji}}=u_{b}^{n-1}|_{\mathcal{K}_{ij}}$ and determine the boundary values of the $\mu_{i}$ from their initial concentrations on $\partial \Omega$.  Since in the barotropic case the total pressure satisfies $p_{b} = \sum_{i}^{\ell}\rho_{i}^{\gamma_{i}}$, we then use the Newton-Raphson method to solve for roots in $\rho_{b}$ of the following equation:
\begin{equation}
\label{acoust}
f(\rho_{b})=\sum_{i}^{\ell}(\rho_{b}\mu_{i,b})^{\gamma_{i}} - (p_{0} + A_{0}\sin(\omega t)).
\end{equation}  This determines the values of $\rho$ on the boundary, where $A_{0}<p_{0}$ is the natural positivity constraint on the pressure inlet.  We allow antisymmetric inlets on $\partial\Omega=\{a,b\}$ leading to the formation of supernodes within the fluid domain.  With our boundary data defined, we utilize the definition: $\boldsymbol{U}_h^{n}|_{K_{ji}}=\{ \rho_b(t^{n}), m_b(t^{n}), \rho_{1,b}(t^{n}), \rho_{2,b}(t^{n}),\rho_{3,b}(t^{n}),\rho_{4,b}(t^{n}),\rho_{5,b}(t^{n}) \}.$  The solution is plotted in Figure \ref{fig:interstellar} for the domain $(0,54)$, a mesh size of $h = 0.135$, a timestep size of $h/30$, and the Runge-Kutta method of order $k=2$. 

By comparison we solve the characteristic acoustic inlet boundary solution using the formalism presented in \textsection{3}.  As in \textsection{4} we linearize about the state $\boldsymbol{U}_{h}^{n}|_{K_{ij}}$ to arrive at an expression for the boundary state $\boldsymbol{U}_{h}^{n}|_{K_{ji}}$ at timestep $t^{n}$.  Now, to determine well-posed characteristic boundary data we must dynamically switch between the five regimes (up to a choice of membrane condition for $u\cdot \boldsymbol{n} =0$) listed in Table \ref{table:odesys}, since the pressure oscillation pulls the velocity between transonic inlet and outlet conditions.  That is, we switch between the following cases:\begin{itemize}
\item Subsonic inlet: $\beta_{1}^{n}$ is fixed by $\boldsymbol{V}^{-1}\tilde{\boldsymbol{q}}_{b}$, while $\beta_{2}^{n},\ldots,\beta_{7}^{n}$ are given by the equations $\mu_{i,b}=\mu_{i,0}$, and $\sum_{i}^{\ell}(\rho_{b}\mu_{i,b})^{\gamma_{i}} - (p_{0} + A_{0}\sin(\omega t))=0$, 
\item Supersonic inlet:  $\beta_{1}^{n},\ldots,\beta_{7}^{n}$ are given by the equations $\mu_{i,b}=\mu_{i,0}$, $u^{n}_{b}=u^{n-1}_{b}$, and $\sum_{i}^{\ell}(\rho_{b}\mu_{i,b})^{\gamma_{i}} - (p_{0} + A_{0}\sin(\omega t))=0$, 
\item Subsonic outlet:  $\beta_{1}^{n},\beta_{3}^{n},\ldots,\beta_{7}^{n}$ are fixed by $\boldsymbol{V}^{-1}\tilde{\boldsymbol{q}}_{b}$, and we solve for $\beta_{2}^{n}$ by way of the pressure equation $\sum_{i}^{\ell}(\rho_{b}\mu_{i,b}(\beta_{2}^{n}))^{\gamma_{i}} - (p_{0} + A_{0}\sin(\omega t))=0$,
\item Supersonic outlet: $\beta_{1}^{n},\beta_{2}^{n},\ldots,\beta_{7}^{n}$ are fixed by $\boldsymbol{V}^{-1}\tilde{\boldsymbol{q}}_{b}$,
\item Wall:  $\beta_{1}^{n},\beta_{3}^{n},\ldots,\beta_{7}^{n}$ are fixed by $\boldsymbol{V}^{-1}\tilde{\boldsymbol{q}}_{b}$, and we solve $\beta_{2}^{n}$ by way of the pressure equation $\sum_{i}^{\ell}(\rho_{b}\mu_{i,b}(\beta_{2}^{n}))^{\gamma_{i}} - (p_{0} + A_{0}\sin(\omega t))=0$,
\end{itemize} where we note that above we have set $\tilde{\boldsymbol{q}}_{b}=\boldsymbol{U}_h^{n}|_{K_{ji}}$.

\begin{figure}[t!]
\centering
\includegraphics[width=10cm]{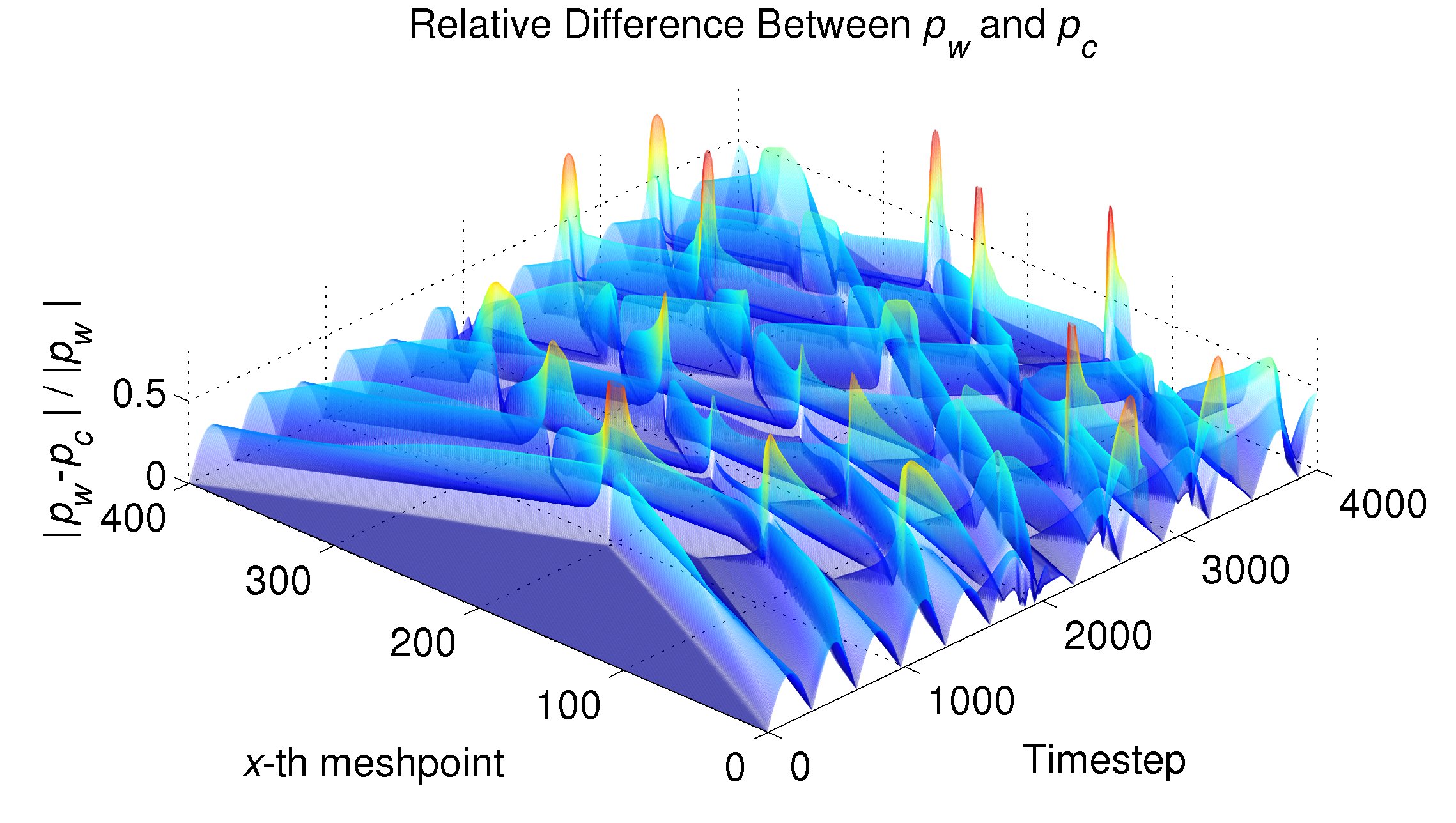}
\caption{Here we plot the relative difference between the weak entropy pressure $p_{w}$ and the characteristic pressure $p_{c}$.}
\label{fig:difference}  
\end{figure}

It can be confirmed by inspection of Figures \ref{fig:interstellar} and \ref{fig:charinterstellar} that the characteristic solution demonstrates substantially sharper profiles than the analogous profiles in the weak entropy solution, and these peaks decay more rapidly in time.  To show this more clearly, we display the difference graph in Figure \ref{fig:difference}.  It is not clear \emph{a priori} which solution is more phenomenologically predictive.

\section{\texorpdfstring{\protect\centering $\S 9$ Conclusion}{\S 9 Conclusion}}

We have shown an efficient and robust high-order numerical scheme for a mixing compressible barotropic viscous fluid comprised of up to $\ell$ distinct chemical constituents.  The DG solution was shown to be in very good agreement with two exact solutions derived by a choice of initial conditions, which demonstrate minimal numerical error at the weak entropy boundaries, as expected.  The solution was then shown for two time-explicit schemes, the forward Euler and $k$-th order explicit Runge-Kutta schemes.  Analysis of the method demonstrated the expected conditional stability up to a restriction by the CFL condition, and we further found that the numerical scheme up to this stability parameter is energy consistent, satisfying a novel entropy inequality; and that the energy consistency hold for a large family of physically relevant problems. We further provide a family of free boundary type solutions which are easily implemented, and which are numerically well-behaved, where either weak entropy or characteristic treatments are employed for comparative studies, and it is seen that indeed they demonstrate distinctly different behaviors even given (seemingly) equivalent initial data. 

A number of examples and potential physical applications were shown and cited in order to develop a sense of the large number of applications in chemistry, physics, engineering, and related fields.   

Future directions of the work include the expansion to higher spatial dimensions ($2$ and $3$ dimensional meshes), the inclusion of Arrhenius type chemical equations to (\ref{specie}), the inclusion of temperature $\vartheta$ dependence into the model, the addition of fluid-structure interfaces, and the expansion of the modelisation to include ionic and polar species as well as dense plasmas (magnetohydrodynamic effects), surface tension and gravitational effects.

\section{\texorpdfstring{\protect\centering $\S$ Acknowledgements}{\S Acknowledgements}}

The first author would like to thank H.~Gupta for providing references on thermophysical properties and additional insights into the interstellar media, and to further express sincere gratitude to Prof.~J.~F.~Stanton for his continued support.  The third author was partially supported by the Department of Energy
Computational Science Graduate Fellowship, provided under grant number
DE-FG02-97ER25308.   The fourth author was partially supported by the NSF Grant DMS 0607953.

\section{\texorpdfstring{\protect\centering $\S$ Appendix}{\S Appendix}}

We have that $\boldsymbol{\Gamma}$ is of the form
\[\boldsymbol{\Gamma}=
\left( \begin{array}{cc|ccc}
   0 & 1  & 0 & \cdots & 0 \\
   -u^2 & 2u & \indet_1 & \cdots & \indet_n \\
   \hline
   -u\mu_1 & \mu_1 \\
   \vdots & \vdots & & u\mathbbm{I}_n \\
   -u\mu_n & \mu_n \\
   \end{array} \right)
\]
where we set for $i=1,\ldots,n$ the indeterminates $\indet_i= \partial_{\rho_i}p$. Solving the characteristic equation $\det\left(\Gamma -\mathbbm{I}\varsigma\right) = 0$,
the eigenvalues counted with multiplicity are,
\[
\varsigma_1=u+c\ ,\varsigma_2=u-c,\quad\varsigma_3=u,\quad\underbrace{\varsigma_4=u,\ldots,\varsigma_{n+2}=u}_{n-1}
\]
where $\displaystyle c = \sqrt{\mu_1\indet_1+\ldots+\mu_n\indet_n}$. While $u$ has multiplicity $n$ it is better to consider the eigenvalues in the three groups, $u\pm c$, $u$, and the remaining $(n-1)$ copies of $u$ as illustrated by the decomposition of the diagonalizing transformation matrix
\[
\boldsymbol{V}(\boldsymbol{U}) = 
(\boldsymbol{c}_1 \cdots \boldsymbol{c}_n) = 
  \left( \begin{array}{ccccccc}
  1      & 1      & 1      & 0       & \cdots & \cdots & 0         \\
  u+c    & u-c    & u      & 0       & \cdots & \cdots & 0         \\
  \mu_1  & \mu_1  & 0      & -\indet_2 & \cdots & \cdots & -\indet_n   \\
  \mu_2  & \mu_2  & 0      & \indet_1  & 0      & \cdots & 0         \\
  \vdots & \vdots & \vdots & 0       & \ddots & \ddots & \vdots    \\
  \vdots & \vdots & \vdots & \vdots  & \ddots & \ddots & 0         \\
  \mu_n  & \mu_n  & 0      & 0       & \ldots & 0      & \indet_1 
  \end{array} \right)
\]
whose columns are the corresponding eigenvectors, which we abbreviate for convenience in the $3 \times 3$ block matrix form
\[
  \left( \begin{array}{cc|c|c}
  1 & 1 & 1 & \mathbf{0} \\
  u+c & u-c & u & \mathbf{0} \\
  \hline
  \mu_1 & \mu_1 & 0 & -\boldsymbol{Y} \\
  \hline
  \boldsymbol{X} & \boldsymbol{X} & \mathbf{0} & (\indet_1)\mathbbm{I}_{n-1}
  \end{array} \right),
\]
where we have set $\boldsymbol{X}=(\mu_2, \ldots, \mu_n)^T$ and
$\boldsymbol{Y}=(\indet_2, \ldots, \indet_n)$.

The inverse transformation matrix is given by
\[
{\renewcommand{\arraystretch}{1.3}%
\boldsymbol{V}^{-1}(\boldsymbol{U}) = \frac{1}{2c^2}
  \left( \begin{array}{cc|c|c}
  -uc & c & \indet_1 & \boldsymbol{Y} \\
  uc & -c & \indet_1 & \boldsymbol{Y} \\
  \hline
  2c^2 & 0 & -2\indet_1 & -2\boldsymbol{Y} \\
  \hline
  \mathbf{0} & \mathbf{0} & -2\boldsymbol{X} & 2\indet_1^{-1}(c^2\mathbbm{I}_{n-1}-\boldsymbol{XY})
  \end{array} \right).
}%
\]

{\setlength\parskip{0pt} 
\bibliography{mybibabrev}{}

\def\cprime{$'$} \def\cprime{$'$}
  \def\polhk#1{\setbox0=\hbox{#1}{\ooalign{\hidewidth
  \lower1.5ex\hbox{`}\hidewidth\crcr\unhbox0}}}
  \def\polhk#1{\setbox0=\hbox{#1}{\ooalign{\hidewidth
  \lower1.5ex\hbox{`}\hidewidth\crcr\unhbox0}}}
  \def\polhk#1{\setbox0=\hbox{#1}{\ooalign{\hidewidth
  \lower1.5ex\hbox{`}\hidewidth\crcr\unhbox0}}}
\begin{thebibliography}{58}
\providecommand{\natexlab}[1]{#1}
\providecommand{\url}[1]{\texttt{#1}}
\expandafter\ifx\csname urlstyle\endcsname\relax
  \providecommand{\doi}[1]{doi: #1}\else
  \providecommand{\doi}{doi: \begingroup \urlstyle{rm}\Url}\fi

\bibitem[Arnold et~al.(2000)Arnold, Brezzi, Cockburn, and Marini]{ABCM}
D.N. Arnold, F.~Brezzi, B.~Cockburn, and D.~Marini.
\newblock Discontinuous {G}alerkin methods for elliptic problems.
\newblock In \emph{Discontinuous Galerkin methods (Newport, RI, 1999)},
  volume~11 of \emph{Lect. Notes Comput. Sci. Eng.}, pages 89--101. Springer,
  Berlin, 2000.

\bibitem[Bardos et~al.(1979)Bardos, le~Roux, and N{\'e}d{\'e}lec]{BLN}
C.~Bardos, A.~Y. le~Roux, and J.-C. N{\'e}d{\'e}lec.
\newblock First order quasilinear equations with boundary conditions.
\newblock \emph{Comm. Partial Differential Equations}, 4\penalty0 (9):\penalty0
  1017--1034, 1979.
\newblock ISSN 0360-5302.

\bibitem[Barth and Charrier(2001)]{Barth2}
T.J. Barth and P.~Charrier.
\newblock Energy stable flux formulas for the discontinuous galerkin
  discretization of first-order conservation laws.
\newblock \emph{Tech. Rep. NAS-01-001, NAS Division, NASA Ames Research
  Center}, 2001.

\bibitem[Bassi and Rebay(1997)]{BR}
F.~Bassi and S.~Rebay.
\newblock A high-order accurate discontinuous finite element method for the
  numerical solution of the compressible {N}avier-{S}tokes equations.
\newblock \emph{J. Comput. Phys.}, 131\penalty0 (2):\penalty0 267--279, 1997.
\newblock ISSN 0021-9991.

\bibitem[Bresch and Desjardins(2007)]{BD3}
D.~Bresch and B.~Desjardins.
\newblock On the existence of global weak solutions to the {N}avier-{S}tokes
  equations for viscous compressible and heat conducting fluids.
\newblock \emph{J. Math. Pures Appl. (9)}, 87\penalty0 (1):\penalty0 57--90,
  2007.
\newblock ISSN 0021-7824.

\bibitem[Bresch and Desjardins(2002)]{BD4}
D.~Bresch and B.~Desjardins.
\newblock Sur un mod\`ele de {S}aint-{V}enant visqueux et sa limite
  quasi-g\'eostrophique.
\newblock \emph{C. R. Math. Acad. Sci. Paris}, 335\penalty0 (12):\penalty0
  1079--1084, 2002.
\newblock ISSN 1631-073X.

\bibitem[Bresch et~al.(2007{\natexlab{a}})Bresch, Desjardins, and
  G{\'e}rard-Varet]{BD7}
D.~Bresch, B.~Desjardins, and D.~G{\'e}rard-Varet.
\newblock On compressible {N}avier-{S}tokes equations with density dependent
  viscosities in bounded domains.
\newblock \emph{J. Math. Pures Appl. (9)}, 87\penalty0 (2):\penalty0 227--235,
  2007{\natexlab{a}}.
\newblock ISSN 0021-7824.

\bibitem[Bresch et~al.(2007{\natexlab{b}})Bresch, Desjardins, and
  M{\'e}tivier]{BD2}
D.~Bresch, B.~Desjardins, and G.~M{\'e}tivier.
\newblock Recent mathematical results and open problems about shallow water
  equations.
\newblock In \emph{Analysis and simulation of fluid dynamics}, Adv. Math. Fluid
  Mech., pages 15--31. Birkh\"auser, Basel, 2007{\natexlab{b}}.

\bibitem[Bridges and Rajagopal(2006)]{BRaja}
C.~Bridges and K.~R. Rajagopal.
\newblock Pulsatile flow of a chemically-reacting nonlinear fluid.
\newblock \emph{Comput. Math. Appl.}, 52\penalty0 (6-7):\penalty0 1131--1144,
  2006.
\newblock ISSN 0898-1221.

\bibitem[Cho et~al.(2004)Cho, Choe, and Kim]{CCK}
Y.~Cho, H.J. Choe, and H.~Kim.
\newblock Unique solvability of the initial boundary value problems for
  compressible viscous fluids.
\newblock \emph{J. Math. Pures Appl. (9)}, 83\penalty0 (2):\penalty0 243--275,
  2004.
\newblock ISSN 0021-7824.

\bibitem[Cockburn(1998)]{Cockburn2}
B.~Cockburn.
\newblock An introduction to the discontinuous {G}alerkin method for
  convection-dominated problems.
\newblock In \emph{Advanced numerical approximation of nonlinear hyperbolic
  equations (Cetraro, 1997)}, volume 1697 of \emph{Lecture Notes in Math.},
  pages 151--268. Springer, Berlin, 1998.

\bibitem[Cockburn and Shu(2001)]{Cockburn1}
B.~Cockburn and C.-W. Shu.
\newblock Runge-{K}utta discontinuous {G}alerkin methods for
  convection-dominated problems.
\newblock \emph{J. Sci. Comput.}, 16\penalty0 (3):\penalty0 173--261, 2001.
\newblock ISSN 0885-7474.

\bibitem[Cockburn and Shu(1998)]{Cockburn3}
B.~Cockburn and C.-W. Shu.
\newblock The local discontinuous {G}alerkin method for time-dependent
  convection-diffusion systems.
\newblock \emph{SIAM J. Numer. Anal.}, 35\penalty0 (6):\penalty0 2440--2463
  (electronic), 1998.
\newblock ISSN 0036-1429.

\bibitem[Cockburn and Shu(1989)]{Cockburn5}
B.~Cockburn and C.-W. Shu.
\newblock T{VB} {R}unge-{K}utta local projection discontinuous {G}alerkin
  finite element method for conservation laws. {II}. {G}eneral framework.
\newblock \emph{Math. Comp.}, 52\penalty0 (186):\penalty0 411--435, 1989.
\newblock ISSN 0025-5718.

\bibitem[Cockburn et~al.(1989)Cockburn, Lin, and Shu]{Cockburn4}
B.~Cockburn, S.-Y. Lin, and C.-W. Shu.
\newblock T{VB} {R}unge-{K}utta local projection discontinuous {G}alerkin
  finite element method for conservation laws. {III}. {O}ne-dimensional
  systems.
\newblock \emph{J. Comput. Phys.}, 84\penalty0 (1):\penalty0 90--113, 1989.
\newblock ISSN 0021-9991.

\bibitem[Dolej{\v{s}}{\'{\i}} and Feistauer(2003)]{FD3}
V.~Dolej{\v{s}}{\'{\i}} and M.~Feistauer.
\newblock On the discontinuous {G}alerkin method for the numerical solution of
  compressible high-speed flow.
\newblock In \emph{Numerical mathematics and advanced applications}, pages
  65--83. Springer Italia, Milan, 2003.

\bibitem[Dolej{\v{s}}{\'{\i}} et~al.(2005)Dolej{\v{s}}{\'{\i}}, Feistauer, and
  Sobot{\'{\i}}kov{\'a}]{FD1}
V.~Dolej{\v{s}}{\'{\i}}, M.~Feistauer, and V.~Sobot{\'{\i}}kov{\'a}.
\newblock Analysis of the discontinuous {G}alerkin method for nonlinear
  convection-diffusion problems.
\newblock \emph{Comput. Methods Appl. Mech. Engrg.}, 194\penalty0
  (25-26):\penalty0 2709--2733, 2005.
\newblock ISSN 0045-7825.

\bibitem[Dukowicz(1980)]{Dukowicz}
J.K. Dukowicz.
\newblock A particle-fluid numerical-model for liquid sprays.
\newblock \emph{J. Comput. Phys.}, 35\penalty0 (2):\penalty0 229--253, 1980.
\newblock ISSN 0021-9991.

\bibitem[Elizarova and Sheretov(2001)]{ES}
T.~G. Elizarova and Yu.~V. Sheretov.
\newblock Theoretical and numerical investigation of quasigasdynamic and
  quasihydrodynamic equations.
\newblock \emph{Comput. Math. Math. Phys.}, 41\penalty0 (2):\penalty0 219--234,
  2001.
\newblock ISSN 0044-4669.

\bibitem[Faeth(1983)]{Faeth}
G.M. Faeth.
\newblock Evaporation and combustion of sprays.
\newblock \emph{Progress in Energy and Combustion Science}, 9\penalty0
  (1--2):\penalty0 1--76, 1983.
\newblock ISSN 0360-1285.

\bibitem[Feistauer and Ku{\v{c}}era(2007)]{FK}
M.~Feistauer and V.~Ku{\v{c}}era.
\newblock On a robust discontinuous {G}alerkin technique for the solution of
  compressible flow.
\newblock \emph{J. Comput. Phys.}, 224\penalty0 (1):\penalty0 208--221, 2007.
\newblock ISSN 0021-9991.

\bibitem[Feistauer et~al.(2003)Feistauer, Felcman, and Stra{\v{s}}kraba]{FFS}
M.~Feistauer, J.~Felcman, and I.~Stra{\v{s}}kraba.
\newblock \emph{Mathematical and computational methods for compressible flow}.
\newblock Numerical mathematics and scientific computation. Oxford University
  Press, 2003.
\newblock ISBN 0-19-850588-4.

\bibitem[Franta et~al.(2005)Franta, M{\'a}lek, and Rajagopal]{FMR}
M.~Franta, J.~M{\'a}lek, and K.~R. Rajagopal.
\newblock On steady flows of fluids with pressure- and shear-dependent
  viscosities.
\newblock \emph{Proc. R. Soc. Lond. Ser. A Math. Phys. Eng. Sci.}, 461\penalty0
  (2055):\penalty0 651--670, 2005.
\newblock ISSN 1364-5021.

\bibitem[Gustafsson and Sundstr{\"o}m(1978)]{GS}
Bertil Gustafsson and Arne Sundstr{\"o}m.
\newblock Incompletely parabolic problems in fluid dynamics.
\newblock \emph{SIAM J. Appl. Math.}, 35\penalty0 (2):\penalty0 343--357, 1978.
\newblock ISSN 0036-1399.

\bibitem[Harlow and Amsden(1975)]{HA}
F.H. Harlow and A.A. Amsden.
\newblock Numerical-calculation of multiphase fluid-flow.
\newblock \emph{J. of Comput. Phys.}, 17\penalty0 (1):\penalty0 19--52, 1975.
\newblock ISSN 0021-9991.

\bibitem[Harrison and van Grieken(1998)]{HG}
R.M. Harrison and R.E. van Grieken.
\newblock \emph{Atmospheric Particles}, volume~5 of \emph{IUPAC Series on
  Analytical and Physical Chemistry of Environmental Systems}.
\newblock John Wiley \& Sons, New York, NY, 1998.
\newblock ISBN 0-471-95935-9.

\bibitem[Kazenkin(2002)]{Kazenkin}
K.~O. Kazenkin.
\newblock Existence of a global generalized solution of a one-dimensional
  problem of the flow of a viscous barotropic gas.
\newblock \emph{Fundam. Prikl. Mat.}, 8\penalty0 (4):\penalty0 993--1007, 2002.
\newblock ISSN 1560-5159.

\bibitem[Kreiss et~al.(1994)Kreiss, Kreiss, and Petersson]{KKP}
G.~Kreiss, H.-O. Kreiss, and N.A. Petersson.
\newblock On the convergence to steady state of solutions of nonlinear
  hyperbolic-parabolic systems.
\newblock \emph{SIAM J. Numer. Anal.}, 31\penalty0 (6):\penalty0 1577--1604,
  1994.
\newblock ISSN 0036-1429.

\bibitem[Kreiss(1970)]{Kriess}
H.-O. Kreiss.
\newblock Initial boundary value problems for hyperbolic systems.
\newblock \emph{Comm. Pure Appl. Math.}, 23:\penalty0 277--298, 1970.
\newblock ISSN 0010-3640.

\bibitem[Lie(2001)]{Lie}
Ivar Lie.
\newblock Well-posed transparent boundary conditions for the shallow water
  equations.
\newblock \emph{Appl. Numer. Math.}, 38\penalty0 (4):\penalty0 445--474, 2001.
\newblock ISSN 0168-9274.

\bibitem[Lions(1998)]{PLL2}
P-L. Lions.
\newblock \emph{Mathematical topics in fluid mechanics. {V}ol. 2}, volume~10 of
  \emph{Oxford Lecture Series in Mathematics and its Applications}.
\newblock The Clarendon Press Oxford University Press, New York, 1998.
\newblock ISBN 0-19-851488-3.
\newblock Compressible models, Oxford Science Publications.

\bibitem[M{\'a}lek and Rajagopal(2007)]{MR2}
J.~M{\'a}lek and K.~R. Rajagopal.
\newblock Incompressible rate type fluids with pressure and shear-rate
  dependent material moduli.
\newblock \emph{Nonlinear Anal. Real World Appl.}, 8\penalty0 (1):\penalty0
  156--164, 2007.
\newblock ISSN 1468-1218.

\bibitem[M{\'a}lek et~al.(2005)M{\'a}lek, Mingione, and Star{\'a}]{MR1}
J.~M{\'a}lek, G.~Mingione, and J.~Star{\'a}.
\newblock Fluids with pressure dependent viscosity: partial regularity of
  steady flows.
\newblock In \emph{EQUADIFF 2003}, pages 380--385. World Sci. Publ.,
  Hackensack, NJ, 2005.

\bibitem[Martin(2007)]{Martin}
S.~Martin.
\newblock First order quasilinear equations with boundary conditions in the
  {$L\sp \infty$} framework.
\newblock \emph{J. Differential Equations}, 236\penalty0 (2):\penalty0
  375--406, 2007.
\newblock ISSN 0022-0396.

\bibitem[Martinell et~al.(2006)Martinell, del Castillo-Negrete, Raga, and
  Williams]{MdRW}
J.J. Martinell, D.~del Castillo-Negrete, A.C. Raga, and D.A. Williams.
\newblock Non-local diffusion and the chemical structure of molecular clouds.
\newblock \emph{Monthly Notices of the Royal Astronomical Society},
  372\penalty0 (1):\penalty0 213--218, 2006.

\bibitem[Matsumura and Nishihara(2001)]{MN}
A.~Matsumura and K.~Nishihara.
\newblock Large-time behaviors of solutions to an inflow problem in the half
  space for a one-dimensional system of compressible viscous gas.
\newblock \emph{Comm. Math. Phys.}, 222\penalty0 (3):\penalty0 449--474, 2001.
\newblock ISSN 0010-3616.

\bibitem[Mellet and Vasseur(2007)]{MV3}
A.~Mellet and A.~Vasseur.
\newblock On the barotropic compressible {N}avier-{S}tokes equations.
\newblock \emph{Comm. Partial Differential Equations}, 32\penalty0
  (1-3):\penalty0 431--452, 2007.
\newblock ISSN 0360-5302.

\bibitem[Michoski and Vasseur(2008)]{MV}
C.~Michoski and A.~Vasseur.
\newblock Existence and uniqueness of strong solutions for a compressible
  multiphase navier-stokes miscible fluid-flow problem in dimension n=1.
\newblock \emph{Math. Models Methods Appl. Sci.}, In Press, 2008.

\bibitem[Moehwald and Shchukin(2006)]{MS}
H.~Moehwald and D.G. Shchukin.
\newblock Sonochemical nanosynthesis at the engineered interface of a
  cavitation microbubble.
\newblock \emph{Physical Chemistry Chemical Physics}, 8\penalty0 (30):\penalty0
  3496--3506, 2006.
\newblock ISSN 1463-9076.

\bibitem[Mucha and Zajaczkowski(2004)]{MW}
P.B Mucha and W.~M. Zajaczkowski.
\newblock Global existence of solutions of the {D}irichlet problem for the
  compressible {N}avier-{S}tokes equations.
\newblock \emph{ZAMM Z. Angew. Math. Mech.}, 84\penalty0 (6):\penalty0
  417--424, 2004.
\newblock ISSN 0044-2267.

\bibitem[Osher(1985)]{Osher}
S.~Osher.
\newblock Convergence of generalized {MUSCL} schemes.
\newblock \emph{SIAM J. Numer. Anal.}, 22\penalty0 (5):\penalty0 947--961,
  1985.
\newblock ISSN 0036-1429.

\bibitem[O'Sullivan and Downes(2006)]{OD}
S.~O'Sullivan and T.P. Downes.
\newblock An explicit scheme for multifluid magnetohydrodynamics.
\newblock \emph{Monthly Notices of the Royal Astronomical Society},
  366\penalty0 (4):\penalty0 1329--1336, 2006.
\newblock ISSN 0035-8711.

\bibitem[Poinsot and Lele(1992)]{PLele}
T.~J. Poinsot and S.~K. Lele.
\newblock Boundary conditions for direct simulations of compressible viscous
  flows.
\newblock \emph{J. Comput. Phys.}, 101\penalty0 (1):\penalty0 104--129, 1992.
\newblock ISSN 0021-9991.

\bibitem[Rawlings and Hartquist(1997)]{RH}
J.M.C. Rawlings and T.W Hartquist.
\newblock Molecular diognostics of diffusive boundary layers.
\newblock \emph{The Astrophysical Journal}, 487:\penalty0 672--688, 1997.

\bibitem[Rudy and Strikwerda(1980)]{RS}
D.H. Rudy and J.C. Strikwerda.
\newblock A nonreflecting outflow boundary condition for subsonic
  {N}avier-{S}tokes calculations.
\newblock \emph{J. Comput. Phys.}, 36\penalty0 (1):\penalty0 55--70, 1980.
\newblock ISSN 0021-9991.

\bibitem[S. et~al.(2001)S., Kuhl, Israelachvili, and Hed]{SIG}
Safran S., T.~Kuhl, J.~Israelachvili, and G.~Hed.
\newblock Polymer induced membrane contraction, phase separation, and fusion
  via marangoni flow.
\newblock \emph{Biophysical Journal}, 81\penalty0 (2):\penalty0 659--666, 2001.
\newblock ISSN 0006-3495.

\bibitem[Shakib et~al.(1991)Shakib, Hughes, and Johan]{Hughes1}
F.~Shakib, T.J.R. Hughes, and Z.~Johan.
\newblock A new finite element formulation for computational fluid dynamics.
  {X}. {T}he compressible {E}uler and {N}avier-{S}tokes equations.
\newblock \emph{Comput. Methods Appl. Mech. Engrg.}, 89\penalty0
  (1-3):\penalty0 141--219, 1991.
\newblock ISSN 0045-7825.
\newblock Second World Congress on Computational Mechanics, Part I (Stuttgart,
  1990).

\bibitem[Solonnikov and Tani(1992)]{ST}
V.~A. Solonnikov and A.~Tani.
\newblock Evolution free boundary problem for equations of motion of viscous
  compressible barotropic liquid.
\newblock In \emph{The Navier-Stokes equations II---theory and numerical
  methods (Oberwolfach, 1991)}, volume 1530 of \emph{Lecture Notes in Math.},
  pages 30--55. Springer, Berlin, 1992.

\bibitem[Solonnikov and Tani(1990)]{ST2}
V.~A. Solonnikov and A.~Tani.
\newblock A problem with a free boundary for {N}avier-{S}tokes equations for a
  compressible fluid in the presence of surface tension.
\newblock \emph{Zap. Nauchn. Sem. Leningrad. Otdel. Mat. Inst. Steklov.
  (LOMI)}, 182\penalty0 (Kraev. Zadachi Mat. Fiz. i Smezh. Voprosy Teor.
  Funktsii. 21):\penalty0 142--148, 173--174, 1990.
\newblock ISSN 0373-2703.

\bibitem[Strikwerda(1977)]{Strikwerda}
J.C. Strikwerda.
\newblock Initial boundary value problems for incompletely parabolic systems.
\newblock \emph{Comm. Pure Appl. Math.}, 30\penalty0 (6):\penalty0 797--822,
  1977.
\newblock ISSN 0010-3640.

\bibitem[Sutherland and Kennedy(2003)]{SK}
J.C. Sutherland and C.A. Kennedy.
\newblock Improved boundary conditions for viscous, reacting, compressible
  flows.
\newblock \emph{J. Comput. Phys.}, 191:\penalty0 502--524, 2003.

\bibitem[Vallis(2006)]{Vallis}
G.~Vallis.
\newblock \emph{Atmospheric and oceanic fluid dynamics : fundamentals and
  large-scale circulation}, volume 2nd Edition.
\newblock Cambridge University Press, New York, NY, 2006.
\newblock ISBN 0-521-84969-1.

\bibitem[van Leer(1997{\natexlab{a}})]{VL1}
B.~van Leer.
\newblock Towards the ultimate conservative difference scheme. {V}. {A}
  second-order sequel to {G}odunov's method [{J}. {C}omput. {P}hys. {\bf 32}
  (1979), no. 1, 101--136].
\newblock \emph{J. Comput. Phys.}, 135\penalty0 (2):\penalty0 227--248,
  1997{\natexlab{a}}.
\newblock ISSN 0021-9991.
\newblock With an introduction by Ch. Hirsch, Commemoration of the 30th
  anniversary \{of J. Comput. Phys.\}.

\bibitem[van Leer(1997{\natexlab{b}})]{VL2}
B.~van Leer.
\newblock Towards the ultimate conservative difference scheme. {IV}. {A} new
  approach to numerical convection.
\newblock \emph{J. Comput. Phys.}, 135\penalty0 (2):\penalty0 227--248,
  1997{\natexlab{b}}.
\newblock ISSN 0021-9991.
\newblock With an introduction by Ch. Hirsch, Commemoration of the 30th
  anniversary \{of J. Comput. Phys.\}.

\bibitem[Williams(1985)]{Williams}
F.A. Williams.
\newblock \emph{Combustion Theory}.
\newblock Combustion Science and Engineering Series. The Benjamin/Cummings
  Publishing Company, Inc., Menlo Park, California, 1985.
\newblock ISBN 0-8053-9801-5.

\bibitem[Youngs(1984)]{Youngs}
D.L. Youngs.
\newblock Numerical-simulation of turbulent mixing by rayleigh-taylor
  instability.
\newblock \emph{Physica D}, 12\penalty0 (1--3):\penalty0 32--44, 1984.
\newblock ISSN 0167-2789.

\bibitem[Zhdanov(2002)]{Zhdanov}
V.M. Zhdanov.
\newblock \emph{Transport Processes in Multicomponent Plasma}.
\newblock CRC, Taylor and Francis, New York, 2002.
\newblock ISBN 0-415-27920-8.

\bibitem[Zlotnik(2008)]{Z2}
A.~A. Zlotnik.
\newblock Parabolicity of a quasihydrodynamic system of equations and the
  stability of its small perturbations.
\newblock \emph{Mat. Zametki}, 83\penalty0 (5):\penalty0 667--682, 2008.

\end{thebibliography}
\bibliographystyle{plainnat}
}

\end{document}